\documentclass[letterpaper]{JHEP3} 
\usepackage{epsfig,array,cite,latexsym,amsmath,oldgerm}
\preprint{INT-PUB 03-14, UW/PT 03-10, BUHEP-03-15}
\newcommand\beq{\begin{equation}}
\newcommand\eeq{\end{equation}}
\newcommand\bal{ \begin{align}}
\newcommand\eal{\end{align} }

\newcommand\eqn[1]{\label{eq:#1}} 
\newcommand\Eq[1]{Eq.~\eqref{eq:#1}} 
\newcommand\half{{\textstyle{\frac{1}{2}}}} 
\newcommand\fourth{{\textstyle{\frac{1}{4}}}}

\newcommand\bfs{\mathbf{S}}
\newcommand\bfz{\mathbf{Z}}
\newcommand\bfl{\boldsymbol{\Lambda}}
\newcommand\bfXi{\boldsymbol{\Xi}}
\newcommand\bfU{\boldsymbol{\Upsilon}}

\newcommand\bftau{\boldsymbol{\tau}}

\newcommand\bfPhi{\boldsymbol{\Phi}}
\newcommand\bfmu{\boldsymbol{\mu}}
\newcommand\bfnu{\boldsymbol{\nu}}

\newcommand{\CC}{{\cal C}}
\newcommand{\CD}{{\cal D}}
\newcommand{\CO}{{\cal O}}
\newcommand{\CN}{{\cal N}}

\newcommand{\CG}{{\cal G}}
\newcommand{\CQ}{{\cal Q}}

\newcommand{\CV}{{\cal V}}
\newcommand{\CW}{{\cal W}}

\newcommand{\rma}{{\textswab{a}}}
\newcommand{\bfn}{{\bf n}}
\newcommand{\bfr}{{\bf r}}

\newcommand{\xh}{\mathbf{\hat{e}}_1}
\newcommand{\yh}{\mathbf{\hat{e}}_2}
\newcommand{\zh}{\mathbf{\hat{e}}_3}
\newcommand{\ih}{\mathbf{\hat{e}}_i}
\newcommand{\jh}{\mathbf{\hat{e}}_j}

\newcommand{\ah}{\mathbf{\hat{e}}_a}
\newcommand{\bh}{\mathbf{\hat{e}}_b}
\newcommand{\ch}{\mathbf{\hat{e}}_c}

\DeclareMathOperator{\Tr}{Tr\,}
\DeclareMathOperator{\Trt}{Tr_2\,}
\newcommand{\sla}[1]%
        {\kern .25em\raise.18ex\hbox{$/$}\kern-.55em #1}
\newcommand{\vev}[1]{\langle #1 \rangle}
\newcommand{\mybar}[1]%
        {\kern 0.6pt\overline{\kern -0.6pt#1\kern -0.6pt}\kern 0.6pt}
\newcommand{\dig}{\kern-1.5pt \raisebox{.9ex}{$\cdot$}  \kern1.5pt
  \raisebox{0ex}{${\mathbf\cdot}$}\kern1.5pt \raisebox{-.9ex}{$\cdot$}} 
\newcommand{\digb}{\kern-1.5pt \raisebox{.75ex}{$\cdot$}  \kern1.5pt
  \raisebox{0ex}{${\mathbf\cdot}$}\kern1.5pt \raisebox{-.75ex}{$\cdot$}} 
\newcommand{\digc}{\kern-1.5pt \raisebox{1.05ex}{$\cdot$}  \kern1.5pt
  \raisebox{0ex}{${\mathbf\cdot}$}\kern1.5pt \raisebox{-1.05ex}{$\cdot$}} 
\newcommand{\drawsquare}[2]{\hbox{%
\rule{#2pt}{#1pt}\hskip-#2pt
\rule{#1pt}{#2pt}\hskip-#1pt
\rule[#1pt]{#1pt}{#2pt}}\rule[#1pt]{#2pt}{#2pt}\hskip-#2pt
\rule{#2pt}{#1pt}}
\newcommand{\Yfund}{\raisebox{-.5pt}{\drawsquare{6.5}{0.4}}}
\newcommand{\Ybarfund}{\mybar{\raisebox{-.5pt}{\drawsquare{6.5}{0.4}}}}%
\newcommand\fverb{\setbox\pippobox=\hbox\bgroup\verb}
\newcommand\fverbdo{\egroup\medskip\noindent%
                        \fbox{\unhbox\pippobox}\ }
\newcommand\fverbit{\egroup\item[\fbox{\unhbox\pippobox}]}
\newbox\pippobox

\title{Supersymmetry on a Euclidean Spacetime Lattice II: \goodbreak Target
  Theories with Eight Supercharges }

\author{Andrew G. Cohen \\ Dept. of Physics, Boston University, 590
  Commonwealth Ave, Boston, MA  02215\\Email: \email{cohen@andy.bu.edu}}

\author{David B. Kaplan \\ Institute for Nuclear Theory, University of
  Washington, 
  Seattle, WA 98195-1550 \\Email: \email{dbkaplan@phys.washington.edu}}

\author{Emanuel Katz \\ Dept. of Physics, University
  of Washington, Seattle, WA 
  98195-1560 \\ Email: \email{amikatz@phys.washington.edu}}

\author{Mithat \"Unsal \\ Institute for Nuclear Theory, University of
  Washington, 
  Seattle, WA 98195-1550 \\Email: \email{mithat@phys.washington.edu}}

\abstract{We formulate Euclidean spacetime lattices whose continuum
  limits are supersymmetric Yang-Mills theories with eight
  supercharges in two and three dimensions.  The lattice actions are
  themselves supersymmetric.}

\begin{document} 

\section{Introduction and results}
\label{sec:1}
A method for constructing lattice regularizations of certain supersymmetric
Yang-Mills (SYM) theories was recently presented in
\cite{Kaplan:2002wv,Cohen:2003aa}.  The basic idea is to create a lattice
which respects a subset of the fermionic symmetries of the SYM target
theory;  these exact symmetries greatly restrict the form of 
relevant operators that can be added to the lattice actions, ensuring 
that the lattice theory flows to the desired target theory in the
continuum limit with little or no fine tuning.  
Realizing this simply stated goal is not so straightforward, however.
Naive implementations of fermionic symmetries in lattice models 
lead to continuum theories which have neither the desired Lorentz
symmetry or supersymmetries.  Our
approach utilizes ``orbifold''
technology introduced originally in string theory
\cite{Douglas:1996sw}, and our lattice 
constructions borrow most directly from work on the deconstruction
of supersymmetric field theories \cite{Arkani-Hamed:2001ca,Arkani-Hamed:2001ie}.
As should be expected, the supersymmetric lattices we construct  do not look very
conventional. Gauge fields appear as noncompact variables, both
complex spin zero
bosons and Dirac spinors have their components dispersed over both
sites and links, and the lattices are not simple cubic
structures. In many cases we are able to show that attaining the
desired continuum limit
of such theories involves no fine tuning of operator
coefficients. Most of these lattices have the fascinating property
that 
nonabelian chiral symmetries result in the continuum limit without fine
tuning, and without having to implement sophisticated fermions, such
as domain wall \cite{Kaplan:1992bt} or overlap
\cite{Narayanan:1995gw,Neuberger:1998fp}.  A serious obstacle to the
numerical implementation of these lattices exists, however, in that
for at least some cases the fermionic determinant is known not to be
positive definite  \cite{Giedt:2003ve}.

In ref. \cite{Kaplan:2002wv} spatial lattices were constructed for a
number of SYM theories with four, eight, and sixteen supercharges, while in
ref. \cite{Cohen:2003aa} the method was applied to the construction
of a four supercharge SYM theory on a two dimensional Euclidean spacetime
lattice.  In this paper we extend this previous work to the study of Euclidean
spacetime lattices for SYM theories with eight supercharges in two
\cite{Diaconescu:1997gu,Witten:1997yu} and
three dimensions \cite{Seiberg:1996bs,Seiberg:1996nz,Chalmers:1997xh}. We are able to show that the two dimensional example
involves no fine 
tuning, while there are two related operators in the three dimensional
case which may
receive logarithmic corrections at one loop (but not at higher
loops). A companion paper describing lattices for sixteen
supercharge SYM theories is in preparation. 


For those uninterested in the technical details of our construction
of supersymmetric lattices, we begin by simply specifying the target theories
we are considering and 
presenting the corresponding lattice actions,
along with the dictionary relating the continuum and lattice variables.
We then proceed to describe, first for the two- and then for the
three-dimensional case,  how we arrive at such lattice actions.  We
make explicit the exact supersymmetries on the lattice using
superfield techniques, and then exploit these symmetries to
understand the continuum limit.
 For
additional discussion of the general method, and for citation of prior
work on supersymmetric lattice theories, we refer the reader to
Refs.~\cite{Kaplan:2002wv,Cohen:2003aa}.

\subsection{ $\CQ=8$ SYM in $d=2$ dimensions}
\label{sec:1a}

The eight supercharge SYM theories we consider can all be obtained by
dimensional reduction of ${\cal N}=1$ SYM in six dimensions.  The two
dimensional version, referred to as $(4,4)$ SYM in $1+1$ Minkowski
dimensions, has a continuum action which can be written in Euclidean
space in terms of two gauge potentials $v_1, v_2$, two Dirac fermions
$\Psi_{\alpha i}$, where $\alpha=1,2$ is the spinor index and $i=1,2$
is a flavor index, and four real scalars $s_\mu=\{s_0, {
  s_a}\}$:

\begin{equation} 
  S= \frac{1}{g_2^2} \int d^2x\, \Tr\Biggl[\frac{1}{4}
  v_{mn} v_{mn} +\frac{1}{2} (D_m s_\mu)^2 +\mybar\Psi_i \gamma_m D_m
  \Psi_i + \mybar \Psi_i [s_0,\,\Psi_i] + i\mybar \Psi_i \gamma_3
  \tau^a_{ij}[{ s_a},\,\Psi_j]
  -\fourth[s_\mu,\,s_\nu]^2\Biggr]\ .
  \eqn{targ2} 
\end{equation} 
(Throughout this paper, indices $\mu,\nu$ run over $0,\ldots,3$,
indices $a,b,c$ run over $1,2,3$, and  indices
$i,j,k$ run over values $1,2$;  the letters $m,n$ are reserved for spacetime indices
running over the values appropriate in the given context. We will use
the letter $\rma$ to refer to the lattice spacing.)
 All fields in the above expression are $k\times k$ matrices,
transforming as adjoints under the $U(k)$ gauge symmetry; $v_{mn}$ is
the gauge field strength, $\bftau$ are the Pauli matrices acting on
the flavor indices, and the three $\gamma$ matrices satisfy the usual
Clifford algebra, $\{\gamma_m,\gamma_n\}=2\delta_{m n}$.  This
action has an $SU(2)^3$ global $R$-symmetry, consisting of the $SU(2)$
$R$-symmetry of ${\cal N}=1$ SYM in six dimensions, and the $SO(4)\sim
SU(2)\times SU(2)$ inherited from the $SO(6)$ Lorentz symmetry of the
six-dimensional theory after dimensional reduction to two dimensions.

The lattice action we propose for simulating this theory is
\begin{equation}
  \begin{aligned}
    S =& \frac{1}{g^2}\sum_{\bfn} \Tr\biggl[\half \left(\mybar z_{i,\bfn
        -\ih} z_{i,\bfn-\ih} - z_{i,\bfn}\mybar z_{i,\bfn}+[\mybar
      z_{3,\bfn} ,\, z_{3,\bfn}]\right)^2\biggr.  \\ &
    +2\left(\left\vert \epsilon_{ij} z_{i,\bfn} z_{j,\bfn + \ih} \right\vert^2+ \left\vert z_{i,\bfn} z_{3,\bfn + \ih} -
        z_{3,\bfn} z_{i,\bfn} \right\vert^2\right) \\
    & +\sqrt{2}\Bigl( \Delta_\bfn(\lambda,\mybar z_a, \psi_a)-
    \Delta_\bfn(\chi,\mybar z_a,\xi_a)
    +\epsilon_{abc}\Delta_\bfn(\psi_a,z_b,\xi_c)\Bigr) \\ &
    +\frac{\rma^2\mu^2}{2}\left[\left( z_{i,\bfn}\mybar
        z_{i,\bfn} -\frac{1}{2\rma^2}\right)^2 +2\frac{\mybar z_{3,\bfn}
        z_{3,\bfn}}{\rma^2}\right] \biggr] \eqn{d2lat}
  \end{aligned} 
\end{equation}
where the sum is over sites $\bfn=\{n_1,n_2\}$ with  $n_{1,2}\in
[1,N]$, with $\xh$, $\yh$ being unit vectors in $n_1$
and $n_2$ directions respectively. 
 All variables are 
$k\times k$ matrices satisfying periodic boundary conditions on the
lattice, and there is an independent $U(k)$ symmetry 
associated with each site, which becomes the $U(k)$ gauge symmetry of
the the continuum theory.   The indices $i,j$
run over $1$ and $2$, the  subscripts $a,b,c$ take on the values
$1,2,3$, and all repeated
indices are summed. In this
expression, the variables $z_a$ and  $\mybar z_a$ refer to complex bosonic variables and their
conjugates, while $\lambda$, $\chi$, $\psi_a$ and $\xi_a$
refer to one-component Grassmann variables.

\begin{figure}[t]
\centerline{
\epsfxsize=6.5cm\epsfbox{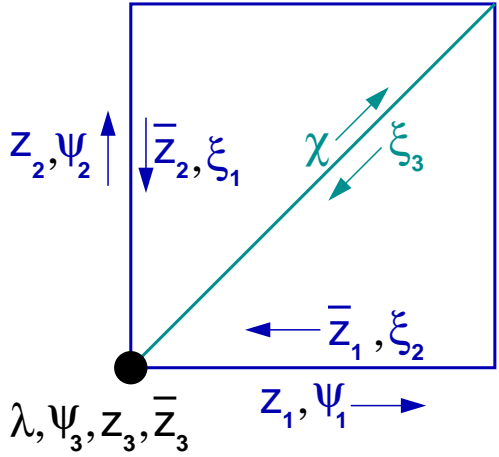}}
\smallskip
\caption{\sl The  unit cell of the Euclidean lattice for the two
  dimensional SYM theory with eight supercharges.  The lattice respects two exact
  supercharges. The figure on the right displays the location of the
  lattice variables associated with site $\bfn$ (Latin for complex bosons,
  Greek for one-component fermions).  Arrows signify the orientation
  of the link variables, or equivalently, the direction of their
  $\mathbf{r}$ charge vectors, as given in Table~1.
  }
  \label{fig:fig1} 
\end{figure}
The structure of the lattice is shown in
Fig.~\ref{fig:fig1}.  We have defined $\Delta$ as
\begin{equation}
  \Delta_{\bfn}(A,B,C) \equiv A_{\bfn} (B_{\bf p}C_{\bf q}-C_{\bf
    r}B_{\bf s})\ , \eqn{deltdef}
\end{equation} 
where the site variables $ {\bf p}$, ${\bf q}$, ${\bf r}$, ${\bf s}$
are determined in terms of $\bfn$ such that each term is gauge
invariant, corresponding to a closed path on the lattice of
Fig.~\ref{fig:fig1}.  For example:
\begin{equation}
  \begin{aligned}
    \Delta_{\bfn}(\lambda,\mybar z_1, \psi_1)&= \lambda_{\bfn}(\mybar
    z_{1,\bfn-\xh}\,\psi_{1,\bfn-\xh}-\psi_{1,\bfn}\,\mybar z_{1,\bfn})\ ,\\
    \Delta_{\bfn}(\chi,\mybar z_3, \xi_3) &= \chi_{\bfn} (\mybar
    z_{3,\bfn+\xh+\yh}\,\xi_{3,\bfn}-\xi_{3,\bfn}\,\mybar z_{3,\bfn})\ 
    ,\\
    \Delta_{\bfn}(\chi,\mybar z_1,\xi_1) &= \chi_\bfn(\mybar z_{1,\bfn +
      \jh} \,\xi_{1,\bfn} -\xi_{1,\bfn+\ih}\, \mybar z_{1,\bfn}).
  \end{aligned}
\end{equation}

The last line in \Eq{d2lat} requires special mention.  The parameter
$\rma$ appearing in the action has dimension of length;  it determines
the expectation values of the link variables $\vev{z_i}$, which in
turn define the  lattice spacing.
  The continuum limit is defined as
$\rma \to 0, N \to \infty$ while holding fixed the 2-dimensional coupling $g_2
\equiv g \rma$ and the lattice size $L \equiv N \rma$.  Special to two
dimensions is that the continuum and thermodynamic limits are not
independent, but must satisfy $g_2\rma \ln N \to 0$ (see
ref. \cite{Cohen:2003aa}). The term
proportional to $\mu^2$ softly breaks the exact supersymmetry, and
controls the size of quantum fluctuations of our dynamical lattice
spacing, $\delta z_i\sim g_2/(\mu L)$. To ensure  that the fluctuations
 $\delta z_i$ are small compared to their mean value ($\delta z_i \ll 
1/\rma$)   $\mu$ must satisfy the constraint $\mu L \gg g_2 \rma $
which can be simply satisfied by taking $\mu \sim 1/L$.

The correspondence between the lattice variables of \Eq{d2lat} and
 the continuum variables in \Eq{targ2} are
\begin{equation}
  \Psi_1 =
  \begin{pmatrix}
    \xi_1\\ \xi_2
  \end{pmatrix}\ , \qquad
  \Psi_2 =
  \begin{pmatrix}
    \ \lambda\\ -\xi_3
  \end{pmatrix} \ ,\qquad
  \mybar\Psi_1 =
  \begin{pmatrix}
    -\chi & \psi_3
  \end{pmatrix} \ ,\qquad
  \mybar\Psi_2 =
  \begin{pmatrix}
    \psi_1 & \psi_2
  \end{pmatrix} \ .
%
  \eqn{ferms}
\end{equation}

\begin{equation}
  \begin{pmatrix}
    s_0\\
    s_1\\
    s_2\\
    s_3
  \end{pmatrix} = 
  \begin{pmatrix}
    \phi_1 \\
    -(z_3+\mybar z_3)/\sqrt{2}\\
    i(z_3-\mybar z_3)/\sqrt{2}\\
    \phi_2
  \end{pmatrix}\ ,\quad \phi_i\equiv \sqrt{2}\,{\rm
    Re}\left(z_i-\frac{1}{\sqrt{2}\,\rma}\right)\ ,\quad
v_m = {\rm Im}\left[z_m\right]\ . 
\eqn{d3dict}
\end{equation}


\subsection{ $\CQ=8$ SYM in $d=3$ dimensions}
\label{sec:1b}

The target theory in three dimensions consists of a gauge potential $v_m$,
where now $m=0,1,2$; two Dirac fermions $\Psi_{\alpha i}$; and three
real scalars $\phi_a$, $a=1,2,3$.  The action respects an $SU(2)\times SU(2)$
global symmetry under which $v_m$ is a $(1,1)$, $\Psi$ and $\mybar\Psi$
together form
a $(2,2)$, and  $\phi_a=(3,1)$.  The action is
\begin{equation}
S = \frac{1}{g_3^2}\int d^3 x\,\Tr\Biggl[\frac{1}{4} v_{mn} v_{mn}
+\frac{1}{2} (D_m \phi_a)^2 +\mybar\Psi_i \sigma_m D_m
\Psi_i  - \mybar \Psi_i 
\tau^a_{ij}\cdot [{ \phi_a},\,\Psi_j]
-\frac{1}{4}[\phi_a,\,\phi_b]^2\Biggr]\ .\\
\eqn{targ3}\end{equation}

\begin{figure}[t]
\centerline{\epsfxsize=7cm\epsfbox{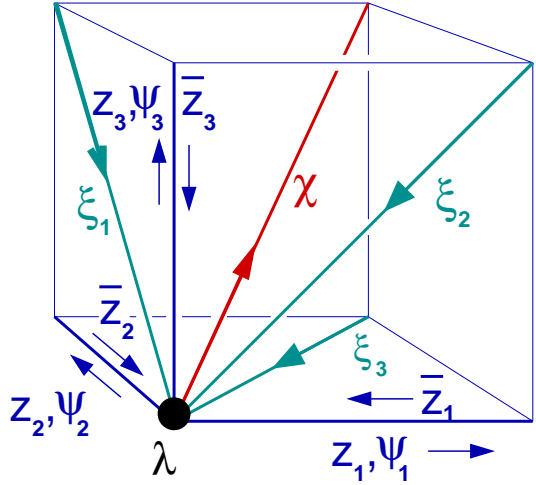}}
\smallskip
\caption{\sl The unit cell of the Euclidean lattice for the three
  dimensional SYM theory with eight supercharges.  Latin and Greek
  letters refer to complex boson and one-component Grassmann
  variables respectively. The lattice respects one exact
  supercharge. }
  \label{fig:fig2} 
\end{figure}

The unit cell for our lattice version of this theory is shown in
Fig.~\ref{fig:fig2}.  The action is given by 
\begin{equation}\begin{aligned} S =&
  \frac{1}{g^2}\sum_{\bfn} \Tr\biggl[\half \left(\mybar z_{a,\bfn-\ah}
    z_{a,\bfn-\ah} - z_{a,\bfn}\mybar z_{a,\bfn}
\right)^2
  +2
\left\vert \epsilon_{abc}\, z_{a,\bfn} z_{b,\bfn + \ah}
\right\vert^2
  \\ & 
+\sqrt{2}\Bigl( \Delta_\bfn(\lambda,\mybar z_a, \psi_a)-
  \Delta_\bfn(\chi,\mybar z_a,\xi_a)
  +\epsilon_{abc}\Delta_\bfn(\psi_a,z_b,\xi_c)\Bigr) 
+\rma^2\mu^2\left( z_{b,\bfn}\mybar z_{b,\bfn}
    -\frac{1}{2\rma^2}\right)^2 
\biggr] 
\\ &
\eqn{d3lat}
\end{aligned} 
\end{equation}
 The
 operator $\Delta$ is defined as in \Eq{deltdef}; however, terms
 differ from the two dimensional case due to the different lattice
 structure.  Now, for example
\begin{equation}
\Delta_{\bfn}(\chi,\mybar z_3, \xi_3) = \chi_{\bfn}
(\mybar z_{3,\bfn+\xh+\yh}\xi_{3,\bfn}-\xi_{3,\bfn+\zh}\mybar z_{3,\bfn})\ ,
\end{equation}
corresponding to the signed sum of the two triangular plaquettes on
 either side for the 
$\chi$ link in the $\chi-\xi_3-\mybar z_3$ plane in Fig.~\ref{fig:fig2}.
The lattice action, with the exception of the soft SUSY breaking terms
proportional to $\mu^2$ which fix the lattice size, respects a single
supercharge.  In the case of this three dimensional lattice, the
continuum limit is defined by $\rma\to 0$, $g\to \infty$, $\rma^3 g^2\to
g_3^2$, $N\to \infty$, keeping $g_3^2$ and $L=\rma N$ fixed.  The
infinite volume limit involves taking $L\to \infty$ and $\mu\to 0$,
with $\mu L$ held fixed and satisfying $\mu L \gg (g_3\,\rma/\sqrt{L})
$; as in the two dimensional case, one can simply fix $\mu L \sim 1$.

In terms of the lattice fields in \Eq{d3lat}, the continuum fields in
\Eq{targ3} are given by
\beq
v_m= \sqrt{2}\, {\rm Im}[z_m]\ ,\qquad \phi_a = \sqrt{2}\, {\rm Re}\left[
  z_a-\frac{1}{\sqrt{2}\,\rma}\right]\ ,
\eeq
for the bosons, and for the fermions
\begin{alignat}{2}
\Psi_1 &= \frac{1}{\sqrt 2}
\begin{pmatrix}
\psi_3-i\chi \cr
 \psi_1+i\psi_2  \\
\end{pmatrix}
\ ,&\qquad
\mybar \Psi_1 &=\frac{i}{\sqrt 2} 
\begin{pmatrix}
\xi_3+i\lambda \cr
 \xi_1-i\xi_2  \\
\end{pmatrix}^T \,\
\\ &&&\\
\Psi_2 &= \frac{1}{\sqrt 2}
\begin{pmatrix}
-\psi_1 + i\psi_2 \cr 
\psi_3+i\chi
\end{pmatrix}\ ,&\qquad
\mybar \Psi_2 &= \frac{i}{\sqrt 2}
\begin{pmatrix}
-\xi_1-i\xi_2  \cr
\xi_3-i\lambda \\
\end{pmatrix}^T\,
\end{alignat}
in a basis where the three gamma matrices are just the Pauli matrices,
\beq
\gamma_m = \sigma_m\ .
\eeq

In the following sections we derive these results and analyze the
renormalization properties of our lattices.

\section{The mother theory and the orbifold projection}
\label{sec:2}

The lattices for $\CQ=8$ SYM theories in two and three dimensions
arise from the $\CQ=8$ mother theory obtained by dimensionally
reducing $\CN=1$ SYM from six (Euclidean) dimensions (a theory
consisting of a six component gauge potential and a four component Weyl
adjoint fermion) to zero dimensions.  The action for the mother theory
may be written as 
\begin{equation} S = \frac{1}{g^2}\left(\frac{1}{4} \Tr v_{mn}
  v_{mn} + \Tr \mybar\psi\, \mybar\Sigma_m [v_m, \psi]\right)\ ,
\eqn{momviiia} 
\end{equation} 
where $m,n=0,\ldots,5$, ${ \psi}$ and $\mybar\psi$
are independent complex four-component spinors and $v_{mn} =
i[v_m,v_n]$.  The variables $\psi$, $\mybar\psi$ and $v_m$ are all
matrices transforming as adjoints under a Lie group $\CG$. The six
$\mybar\Sigma_m$ matrices are defined as \begin{equation} \mybar\Sigma_m =
\{1,i{\boldsymbol\gamma}\}=\Sigma_m^\dagger \eqn{gam}\end{equation} where
${\boldsymbol\gamma}$ are the five four-dimensional gamma matrices
obeying the Clifford algebra for $SO(5)$.

The global symmetry of the action  \Eq{momviiia} is $G_R=SO(6)\times
SU(2)$.  The $SO(6)$ symmetry is just the (Euclidean version of the)
inherited six dimensional Lorentz symmetry, but the form of the action
\Eq{momviiia} makes the $SU(2)$ symmetry far from manifest.  We
therefore define $C$ to be the charge conjugation matrix for $SO(5)$
satisfying
\begin{equation}
C=C^\dagger=C^{-1}=-C^T\ ,\quad C{\boldsymbol \gamma}C = +{\boldsymbol \gamma}^T\
,
\end{equation}
and define the $4\times 2$ Grassmann field
\begin{equation}
\Psi = 
\begin{pmatrix}
\psi\ \ \  & C \mybar\psi^{\,T}
\end{pmatrix}\ .
\end{equation}
In the above expression, the transpose in $\mybar\psi^{\,T}$ affects the
Dirac indices only, and not the gauge indices.  Then in terms of
$\Psi$, the action \Eq{momviiia} may be reexpressed as 
\begin{equation}
 \frac{1}{g^2}\left(\frac{1}{4} \Tr v_{mn} v_{mn} -\frac{i}{2}\Tr \sigma_2\, \Psi^T C\,\mybar \Sigma_m \left[v_m,\,
  \Psi\right]\right)\ .
\eqn{momviii}
\end{equation}
It is now evident that this mother theory has a global $G_R=SO(6)\times SU(2)$ symmetry
under which 
\begin{equation}
v\to \Omega v \Omega^{-1}\ ,\qquad \Psi\to \Omega \Psi
U^\dagger\ ,\qquad \text{with} \quad v\equiv
C\mybar\Sigma_m v_m \ ,
\eqn{uom}\end{equation}
where $\Omega\in SO(6)$ and $U\in SU(2)$.
 
The eight supersymmetry transformations may be parametrized by a
constant Grassmann $4\times 2$ dimensional spinor $\kappa$.  The
action \Eq{momviii} is invariant under the supersymmetry
transformations
\begin{equation} \delta v_m = \Trt \sigma_2 \,\kappa^T C
 \,\mybar\Sigma_m \Psi\ ,\quad \delta\Psi = -i v_{m n} \Sigma_{m n}
 \kappa\ ,\quad \delta\Psi^T = -i v_{m n}\kappa^T \Sigma_{m n}^T \ ,
 \eqn{dq6} 
\end{equation} 
where $\Trt$ means a trace over the $2\times 2$
 matrix, and not over gauge indices, and we have defined the matrices
\begin{equation} \Sigma_{m n} = \frac{i}{4}\left(\Sigma_m\mybar\Sigma_ n -
   \Sigma_ n\mybar\Sigma_m\right)\ .  \eqn{sigdef}
\end{equation}

A lattice may be created out of the matrices of the mother theory,
following the procedure in \cite{Cohen:2003aa}. To create a
$d$-dimensional lattice with $N^d$ sites and possessing a $U(k)$ gauge
symmetry, we take the group $\CG$ of the mother theory to be
$U(k N^d)$.  The variables of the mother theory are then all
$kN^d\times kN^d$ matrices. A given matrix variable $\Phi$ of the
mother theory is most conveniently labeled not with two indices, but
as $\Phi^{(\bfmu,\bfnu)}_{ij}$, where $i$, $j$ run from $1,\ldots,k$,
while $\bfmu$ and $\bfnu$ are $d$-component vectors, with each element
running over $1,\ldots,N$.  We then define a particular $\Gamma=Z_N^d$
subgroup of the $G_R\times \CG$ symmetry of the mother theory, whose
generators $\hat \gamma_a$ act on the field $\Phi$ as
\begin{equation}
  \hat{\gamma}_a \Phi = e^{2\pi i r_a/N} \CC^{(a)} \,\Phi\, {\CC^{(a)}}^{-1}\
  \qquad \text{with}\quad a=1,\ldots,d\ .\ 
  \eqn{orbi}
\end{equation}
In this expression,  $e^{2\pi i r_a/N}\in G_R$ and $\CC^{(a)}\in \CG$.  The
$\CC^{(a)}$ are referred to as ``clock'' matrices. They can be written
compactly
as the direct product of $d$ rank-$N$ matrices and one rank-$k$ matrix:
\begin{equation}
\begin{aligned}
  \CC^{(1)} &= \Omega\otimes {\mathbf 1}_N \otimes \cdots\otimes
  {\mathbf 1}_N\otimes {\mathbf 1}_k\ ,\\ 
  \CC^{(2)} &=  {\mathbf 1}_N\otimes \Omega \otimes  \cdots\otimes 
  {\mathbf 1}_N\otimes{\mathbf 1}_k\ ,\\
\ldots &&\\
  \CC^{(d)} &=  {\mathbf 1}_N\otimes {\mathbf 1}_N\otimes\cdots\otimes \Omega \otimes
 {\mathbf 1}_k\ ,
\end{aligned}
\eqn{clocka}
\end{equation}
where ${\mathbf 1}_m$ signifies a rank-$m$ unit matrix, and $\Omega$ is
the rank-$N$ unitary matrix
\begin{equation}
  \Omega = 
  \begin{pmatrix} 
    \  \, \omega\  \,  & & &\\ &\  \,
    \omega^2\,  \ & & \\ &&\dig &\\
    &&&\omega^{N} 
  \end{pmatrix}
  \qquad \text{with} \quad \omega\equiv e^{2\pi i/N}\ .
\eqn{clockb}\end{equation}
When multiplied on the left of
$\Phi^{(\bfmu,\bfnu)}_{ij}$ $\CC^{(m)}$ acts nontrivially only on the
$m^{th}$ component of the index vector $\bfmu$, while
${\CC^{(m)}}^{-1}$ acts nontrivially only on the 
$m^{th}$ component of the index  vector $\bfnu$ when multiplied on the
right.

The integer charges $\mathbf{r}=\{r_1,\ldots,r_d\}$ in \Eq{orbi} are
constructed from the Cartan sub-algebra of $G_R= SO(6)\times
SU(2)$. Our specific choice of these charges is discussed below. We
will  choose a basis for the variables of the mother theory 
such that each $\Phi^{(\bfmu,\bfnu)}_{ij}$ is an eigenstate of these
$d$ charges, and so each factor $e^{2\pi i r_a/N}$ in \Eq{orbi} will
be a simple phase, and will act trivially on the indices of $\Phi$

The orbifold projection eliminates the components of the mother theory
fields which are not invariant under the discrete $\Gamma$ transformation
defined in \Eq{orbi}.  The projection operator may be written as
\begin{equation}
  \hat{P} = \frac{1}{N^d}\sum_{m_1=1}^N\ldots\sum_{m_d=1}^N\
  (\hat{\gamma}_1)^{m_1}\cdots(\hat{\gamma}_d)^{m_d}\ .
\end{equation}
The ``daughter'' theory is obtained by replacing every field $\Phi$ in
the action of the mother theory by its projection $\tilde{\Phi} =
\hat{P} \Phi$.  Each projected field $\tilde{\Phi}$ is very sparse,
consisting only of $N^d$
nonzero (unconstrained) $k\times k$ blocks; the position of these nonzero blocks
within the original rank $k\,N^d$ matrix is determined by the $\bfr$
charges of $\Phi$. In particular, the
nonzero blocks in $\tilde\Phi^{(\bfmu,\bfnu)}_{ij}$ occur only for
\begin{equation}
  \bfnu = \bfmu + \bfr\ .
  \eqn{rmean}
\end{equation}
Recall that $\bfmu$, $\bfnu$ and $\bfr$ are each $d$-component vectors
with integer components running from $1$ to $N$.  We can therefore
consider these nonzero $k\times k$ blocks as lattice variables of a
$d$-dimensional, $N^d$-site lattice.  We label each lattice site by a
vector $\bfmu$.  Each nontrivial block in $\tilde \Phi$ is then a
variable residing on the link between sites $\bfmu$ and $\bfmu+\bfr$.
Since the orbifold projection breaks the symmetry $\CG$ of the mother
theory down to an independent $U(k)$ symmetry for each of the $N^d$
sites of the lattice, a link variable between sites $\bfmu$ and
$\bfmu+\bfr$ transforms under the bilinear $(\,\Yfund,\Ybarfund\,)$
representation of the $U(k)\times U(k)$ symmetry associated with sites
$\bfmu$ and $\bfmu+\bfr$. A variable of the mother theory with
$\bfr=0$ corresponds to site variables on the lattice, each one
transforming under the adjoint representation of the $U(k)$ symmetry
associated with that site.

In order to make this explicit, we must choose the $\bfr$ charges, and
express the variables of the mother theory in terms of eigenstates of
these charges.  We begin by defining a basis $q_1,\ldots,q_4$ for
the Cartan sub-algebra of the rank 4 group $G_R= SO(6)\times SU(2)$.
The generators of the $SO(6)$ are just the matrices
$\Sigma_{mn}$ defined in \Eq{sigdef}, while we denote the three generators
of the $SU(2)$ as $\half {\boldsymbol \tau}$.  We choose as our
basis the four mutually commuting charges
\begin{equation}
q_1 =\Sigma_{01}
\ ,\quad q_2=\Sigma_{23}
\ ,\quad q_3= \Sigma_{45}
\ ,\quad q_4=\half \tau_3\ .
\eqn{qs}
\end{equation}

Next we  define the three bosonic variables $z_a$ and their Hermitean
conjugates $\mybar z_a$,  which are linear
combinations of the six gauge fields $v_m$ of the
mother theory, and which are eigenstates of the
$q_i$: 
\begin{equation}
\begin{array}{rclcrclcrcl}
 z_1 &=& +i\frac{v_0+iv_1}{\sqrt{2}}\ ,&\quad&
 z_2&=& \frac{v_2+iv_3}{\sqrt{2}}\ ,&\quad&
 z_3 &=& \frac{v_4+iv_5}{\sqrt{2}}\ ,\\
&&&&&&&&&&\\
\mybar z_1 &=& -i\frac{v_0-iv_1}{\sqrt{2}}\ ,&\quad&
\mybar  z_2&=& \frac{v_2-iv_3}{\sqrt{2}}\ ,&\quad&
\mybar z_3 &=& \frac{v_4-iv_5}{\sqrt{2}}\ .
\end{array}
\end{equation}
The $q_i$ charges of these variables are given in Table~\ref{tab:tab1}.

At this point it is convenient to use the particular basis for the
$SO(5)$ gamma matrices of \Eq{gam}
\begin{equation}
\begin{array}{rclcrclcrcl}
\gamma_1&=&-\sigma_3\otimes 1\ ,&\quad& 
\gamma_2&=&\sigma_1\otimes \sigma_1\ ,&\quad&
\gamma_3&=&-\sigma_1\otimes \sigma_2\ ,\\
\gamma_4&=&-\sigma_1\otimes \sigma_3\ ,&\quad&
\gamma_5&=&\sigma_2\otimes 1\ ,&\quad& C &=& \sigma_3\otimes\sigma_2\
.
\end{array}
\end{equation}
In this basis  the $q_i$ are diagonal,
\begin{equation}
q_1= \half\, \sigma_3\otimes 1
\ ,\quad q_2= \half\,1\otimes\sigma_3
\ ,\quad q_3=\half\,\sigma_3\otimes\sigma_3
\ ,\quad q_4=\half \tau_3\ ,
\eqn{qsb}
\end{equation}
and the gauge potential matrix takes the particularly
simple form
\begin{equation}
v = C\mybar\Sigma_m v_m=\sqrt{2}\,
\begin{pmatrix}
 \ \ 0 &  \ \ \mybar z_1 &  \ \ \mybar z_2& \ \ \mybar z_3 \\
-\mybar z_1 & \ \  0 &  \ \  z_3  & -z_2 \\ 
-\mybar z_2& - z_3  & \ \  0 &\ \  z_1 \\
 -\mybar z_3 &  \ \ z_2 & -z_1  &\ \  0
\end{pmatrix}
\ .\end{equation} 
Furthermore, the individual components of $\Psi$ are $q_i$ eigenstates
in this basis.  We label them as
\begin{equation}
\Psi = 
\begin{pmatrix}
\lambda & \chi \\ 
\xi_1 & \psi_1\\
\xi_2 & \psi_2 \\ 
\xi_3 & \psi_3
\end{pmatrix}\ ,
\end{equation}
and give their $q_i$ charges in Table~\ref{tab:tab1} as well.
In terms of these variables, the action of the mother
theory \Eq{momviiia} becomes
\begin{equation}
S=
\frac{1}{g^2}\Tr\Biggl[
\half\, [\mybar z_a, z_a ]^2
+2 \bigl|\epsilon_{abc} z_a z_b \bigr|^2
+
\sqrt{2}\left(\lambda\,[\mybar z_a,\psi_a] - \chi\,
  [\mybar z_a,\xi_a] + \epsilon_{abc}\psi_a\,[ z_b,\xi_c]\right)\Biggr]\
.
\eqn{act8}
\end{equation}

 \setlength{\extrarowheight}{5pt}
\begin{table}[t]
\centerline{
\begin{tabular}
{|c||c|c|c||c|c|c|c||c|c|c|c|}
\hline
&$  z_1$&$ z_2$&$ z_3$
&$ \lambda$
&$\xi_1$&$\xi_2$&$\xi_3$
&$ \chi$
&$\psi_1$&$\psi_2$&$\psi_3$ \\ \hline
$q_1 $&$
+1 $&$ \,\ 0 $&$ \,\ 0  $&$ +\half $&$ +\half $&$ -\half $&$ -\half $&$ +\half $&$ +\half $&$ -\half $&$ -\half$ \\ 
$q_2 $&$ 
\,\ 0  $&$  +1 $&$  \,\ 0   $&$  +\half $&$  -\half $&$  +\half $&$  -\half
$&$  +\half  $&$ -\half $&$  +\half $&$  -\half$ \\ 
$q_3 $&$ 
\,\ 0 $&$ \,\ 0 $&$ +1  $&$ +\half $&$ -\half $&$ -\half $&$ +\half $&$ +\half
$&$ -\half $&$ -\half $&$ +\half$ \\ 
$q_4 $&$ 
\,\ 0$&$ \,\ 0$&$ \,\ 0$&$  -\half$&$ -\half$&$ -\half$&$ -\half$&$ +\half$&$
+\half$&$ +\half$&$ +\half$ \\   \hline
$r_1 $&$ 
+1 $&$  \,\ 0 $&$  \,\ 0 $&$  \,\ 0 $&$  \,\ 0 $&$  -1 $&$  -1 $&$  +1 $&$  +1 $&$  \,\ 0
$&$  \,\ 0$ \\ 
$r_2 $&$ 
\,\ 0 $&$ +1 $&$ \,\ 0 $&$ \,\ 0 $&$ -1 $&$ \,\ 0 $&$ -1 $&$ +1 $&$ \,\ 0 $&$ +1 $&$ \,\ 0 $ \\ 
$r_3 $&$ 
\,\ 0$&$ \,\ 0$&$ +1$&$ \,\ 0$&$ -1$&$ -1$&$ \,\ 0$&$ +1$&$ \,\ 0$&$ \,\ 0$&$ +1 $
\\ \hline
 \end{tabular}
}
\caption{\sl The  $q_{1,\ldots,4}$ charges of the boson, fermion, and
  auxiliary fields of the $\CQ=8$ mother 
 theory under the $U(1)^4$ subgroup of $G_R=SO(6)\times SU(2)$.   The
  $r_{1,2,3}$ charges are the linear combinations of the $q$ charges
which
  define the possible orbifold projections. The $\mybar z_a$ fields
  have  charges opposite to those of their unbarred 
  counterparts.\label{tab:tab1}} 
\end{table}
 \setlength{\extrarowheight}{-2pt}

The next task is to construct the $\bfr$ charges out of independent
linear combinations of the  $q_i$.
The criterion for choosing one combination over the other is that (i)
all $r_a$ charges must be integer for \Eq{orbi} to define a $Z_N$
transformation; (ii) all $r_a$ charges should be $0$ or $\pm1$ if we only
want interactions between neighboring sites of the lattice; (iii) as
explained in Refs.~\cite{Cohen:2003aa,Kaplan:2002wv}, the number of unbroken
supersymmetries equals the number of $\bfr=0$ Grassmann variables, so
we want the maximum number of fermion components (half of them) to have
$r_a=0$ for each $a$.  These considerations lead us to define
the three $r_a$ charges:
\begin{equation}
r_1 = q_1+q_4\ ,\qquad r_2 = q_2 +q_4\ ,\qquad r_3=q_3+q_4\ .
\eqn{rs}
\end{equation}
Table~\ref{tab:tab1} lists the $\bfr$ charges for each of the
variables in the mother theory.

By choosing the orbifold group $\Gamma$ to be $Z_N$,
$Z_N^2$ or $Z_N^3$ we will obtain one-, two-, or three-dimensional lattices,
possessing four, two or one supersymmetries respectively.  We could in
principle create a four dimensional lattice, but it would not respect
any exact supersymmetry.  In the remainder of this article we focus on
the two- and three-dimensional lattices.

\section{The two dimensional lattice}
\label{sec:3}

To create a two dimensional lattice from the $\CQ=8$ mother theory,
which will describe the $(4,4)$ SYM theory in the two-dimensional
continuum, we orbifold by $Z_N\times Z_N$, where the two $Z_N$
transformations are determined by the charges $r_1$ and $r_2$ in
Table~\ref{tab:tab1}.  The lattice we obtain takes the form shown in
Figure~\ref{fig:fig1}, with $\{z_3, \mybar z_3, \lambda\}$ residing at the
sites; $\{z_1,\mybar z_1, \xi_2, \psi_1\}$ on the $\xh$-links, $\{z_2,\mybar
z_2, \xi_1, \psi_2\}$ on the $\yh$-links; and $\{\xi_3, \psi_3\}$ along
the diagonal links.  The lattice action in terms of the component
fields follows immediately from our ${\bf r}$ charge assignments and
\Eq{act8}, and is the action of \Eq{d2lat}, with the omission of the
soft supersymmetry breaking term proportional to $\mu$.  The
symmetries of this action include $\CQ=2$ supersymmetry, the $U(k)$
gauge symmetry, a $U(1)^4$ global symmetry (generated by the four
$q_i$ charges), and a $C_{2v}$ lattice symmetry. The generators of the
$C_{2v}$ symmetry consist of reflections about the diagonal axis,
$\sigma_d$, and $\pi$ rotations about the normal to the lattice,
$C_2$.  It is worthwhile to make these symmetries manifest so that we
can more easily analyze the approach to the continuum.

\subsection{The $\CQ=2$ supersymmetry of the $d=2$ lattice}
\label{sec:3a}

We begin by rewriting the action in a superfield
formalism, which will make  the $\CQ=2$ symmetry manifest.

To find the exact supersymmetry transformations, we need only find the
subset of the full supersymmetry transformations in \Eq{dq6} which
commute with the ${\bf r}=\{r_1,r_2\}$ charges.  One finds these to
correspond to restricting the supersymmetric parameter $\kappa$ in
\Eq{dq6} to the form 
\begin{equation} 
\kappa = 
\begin{pmatrix}\ \eta\ & \ 0\ \\ \ 
  0\ & \ 0\ \\ \ 0\ & \ 0\ \\ \ 0\ & \ \mybar \eta
\end{pmatrix}
\end{equation} 
where $\eta$ and $\mybar \eta$ are independent one-component
Grassman parameters.  Then the supersymmetry transformations in the
mother theory in \Eq{dq6}, restricted to $\kappa$ as given above, are
\begin{eqnarray}
\begin{array}{rclcrcl}

\delta z_i &=& i\sqrt{2}\,\eta\,\psi_i &\qquad& \delta\mybar z_i
&=&i\epsilon_{ij}\,\sqrt{2}\,\mybar\eta\,\xi_j\\
\delta\psi_i&=&2i\mybar\eta\,[z_i,\,\mybar z_3]  &\qquad&
\delta\xi_i&=& -2i\epsilon_{ij}\,\eta\,[\mybar z_j,\,\mybar z_3]\\  &&&&&&\\
%
\delta z_3 &=& i\sqrt{2}\,(\eta\,\psi_3 +\mybar \eta\, \lambda)&\qquad& \delta\mybar z_3
&=&0\\
\delta\psi_3&=&i\mybar\eta\,\left([\mybar z_i,z_i]
 - [\mybar z_3,\,z_3]\right)
&\qquad\qquad  &
\delta\lambda &=&-i\eta\,\left([\mybar z_i,\,z_i]
 
+ [\mybar
  z_3,\,z_3]\right) \\  &&&&&&\\
\delta\chi&=&2i \mybar\eta\,[z_1,\,z_2]\mybar
 &\qquad& \delta\xi_3 &=&-2i\eta\,[\mybar z_1,\,\mybar z_2] \ .
\end{array}
\eqn{dfields}
\end{eqnarray}
The fact that both sides of these transformations carry the same
$\bfr$ charges is equivalent to the statement that the orbifold
projection leaves these supersymmetry transformations unbroken. We
have grouped together fields according to their $\bfr$ charges: the
three groups above reside on the lattice at the $x$- and $y$-links,
sites, and diagonal links respectively.

The
supersymmetry transformations \Eq{dfields}  may be written  in terms
of two supercharges $Q$ and 
$\mybar Q$ as
\begin{equation}
\delta = i\eta Q + i \mybar \eta\mybar Q\ .
\end{equation}
The  $Q$
and $\mybar Q$ supercharges may be realized in terms of the
independent Grassmann coordinates 
$\theta$ and $\mybar \theta$ by defining
\begin{equation}
Q =\frac{\partial\  }{\partial \theta} + \sqrt{2}\, \mybar\theta[\mybar z_3,\,\cdot \ ] \ ,\qquad
\mybar Q =\frac{\partial\  }{\partial \mybar\theta} +   \sqrt{2}\, \theta[\mybar z_3,\,\cdot \ ] \ .
\end{equation}
This requires the introduction of three auxiliary fields $d$,
$G$ and $\mybar G$, since the last two lines of \Eq{dfields} are only
compatible with $Q^2 = \mybar Q^2 =0$ after invoking the equations of
motion.  The auxiliary fields modify the variations of $\lambda$, $\chi$,
$\psi_3$, and $\xi_3$ which now read:
 \begin{eqnarray}
\begin{array}{rclcrcl}
\delta\psi_3 &=& i\mybar\eta\,\left([\mybar z_i,z_i]
- [\mybar
  z_3,z_3]-id\right) 
&\qquad\qquad  &
\delta\lambda &=&-i\eta\,\left([\mybar z_i,z_i]
+ [\mybar   z_3,z_3]+id\right) \\ &&&&&& \\
\delta\chi&=&i\mybar \eta\,(2\,[z_1,z_2]-\sqrt{2}\,\mybar G)
 &\qquad& \delta\xi_3 &=&-i\eta\,(2\,[\mybar z_1,\mybar
 z_2]-\sqrt{2}\,G) 
\end{array}
\end{eqnarray}
while the auxiliary fields vary in such a way as to ensure $Q^2=\mybar
Q^2=0$:
\begin{equation}\begin{aligned}
\delta \mybar G &= 2i\eta\, \epsilon_{ij}\,[z_i,\psi_j]\, 
\\  
\delta G &= -2i\mybar\eta \,[z_i,\xi_i]\\
\delta d &= -\sqrt{2}\,\eta\, 
\left([\mybar z_i,\psi_i]
+[\mybar z_3,\psi_3]\right)-i\sqrt{2}\,\mybar\eta\left(\epsilon_{ij}\,[z_i,\xi_j]+[\mybar
  z_3,\lambda]\right)\ .
\end{aligned}\end{equation}

We can also define supersymmetric derivatives which anticommute with
the $Q$'s:
\begin{equation}
{\CD}=\frac{\partial\  }{\partial \theta} - \sqrt{2}\, \mybar\theta\,[\mybar z_3,\,\cdot \ ] \ ,\qquad
{\mybar\CD} =\frac{\partial\  }{\partial \mybar\theta} -  \sqrt{2}\, \theta\,[\mybar z_3,\,\cdot \ ] \ .
\end{equation}
All of the link fields  may then be combined
into chiral superfields:
\begin{equation}
\begin{aligned}
{\bfz_i} &= z_i +\sqrt{2}\,\theta \psi_i -\sqrt{2}\, \theta\mybar\theta
[\mybar z_3,z_i]\ ,
&\mybar {\bfz}_i &= \mybar z_i
+\sqrt{2}\,\mybar\theta \epsilon_{ij}\, \xi_j +\sqrt{2}\, \theta\mybar\theta [\mybar z_3,\mybar z_i]\ , \\
%
{\mathbf \Xi} &= \xi_3 +\sqrt{2}\, \theta (G-\sqrt{2}\,[\mybar
z_1,\mybar z_2])  -\sqrt{2}\, \theta \mybar\theta 
[\mybar z_3,\xi_3]\ ,\quad 
&\mybar{\bold \Xi} &= \chi -\sqrt{2}\,\mybar\theta(\mybar
G-\sqrt{2}\,[ z_1, z_2])   +\sqrt{2}\,\, \theta \mybar\theta 
[\mybar z_3,\chi]\ .
\end{aligned}\end{equation}
satisfying the chiral constraints  ${\mybar\CD}{\bfz_i}={\mybar\CD}{\bf \Xi}=0$, and
 ${\CD}\mybar {\bfz}_i= {\CD} \mybar {\bold
  \Xi}=0$. 
The site variables reside in a non-chiral superfield
\begin{equation}
{\bfs} = z_3 + \sqrt{2}\,\theta \psi_3 + \sqrt{2}\, \mybar\theta \lambda
+\sqrt{2}\,\theta \mybar\theta ([\mybar z_i,z_i] +id)\ .
\end{equation}
From the ${\bfs}$ superfield we can create the chiral and anti-chiral
superfields which will appear in the action:
\begin{equation}
\begin{aligned}
 \mybar {\bold\Upsilon}&= \frac{{\CD}{\bfs}}{\sqrt2}= \psi_3 + \mybar{\theta}(
[\mybar{z}_i, z_i]-[\mybar{z}_3, z_3] +i d) 
+ \sqrt2 \theta \mybar{\theta}[\mybar{z}_3,\psi_3]   , \\
{\bold \Upsilon}&= \frac{\mybar{{\CD}}{\bfs}}{\sqrt2}= \lambda - 
\theta ([\mybar{z}_i, z_i]+[\mybar{z}_3, z_3] +i d)
-\sqrt2 \theta \mybar{\theta}[\mybar{z}_3, \lambda] \ .
\end{aligned}\end{equation}

Note that $\mybar z_3$ is a singlet under the two supersymmetries, and
so is a superfield all by itself. Also note that the $\theta$ ($\mybar
\theta$) components of (anti-)chiral superfields, and the
$\mybar\theta\theta$ components of a general  superfield, transform under
supersymmetry into a commutator;  therefore the trace of such terms
are supersymmetric invariants and are suitable for construction of the
action.

In terms of these superfields, the action of the mother
theory in \Eq{act8} may be written as
\begin{equation}\begin{aligned}
S &= \int d\theta d\mybar\theta\, \Tr\left(
    \frac{1}{2}\mybar{\bold\Upsilon}{\bold\Upsilon} + \frac{1}{\sqrt2} 
{\mybar\bfz}_i 
[{\bfs},{\bfz_i}]
- \frac{1}{2}\mybar{\bold\Xi} 
{\bold \Xi}\right)
\\ &\qquad +\int d\theta \, \Tr\left({\bold \Xi}\,[{\bfz_1},{\bfz_2
  }]\right)
 -\int d\mybar\theta \, \Tr\left(\mybar{\bold\Xi}\,[\mybar {\bfz}_1, 
{\mybar\bfz}_2]\right)
\ ,
\end{aligned}\eqn{act8q2}\end{equation}
  The only
difference between the above action and that of \Eq{act8} is the
addition of the auxiliary variables $d$, $G$ and $\mybar G$, which only
enter the action as
\begin{equation}
 S_{\rm aux} = \Tr\left[\frac{d^2}{2} + \mybar G G\right]\ ,
\end{equation}
with the equations of motion $d=G=\mybar G=0$, and have no
effect on the dynamics of the theory.  

After performing the orbifold projection, 
the superfields take the form
\begin{equation}
\begin{aligned}
{\bfz_{i,\bfn}} &= z_{i,\bfn} +\sqrt{2}\,\theta \psi_{i,\bfn} -\sqrt{2}\, \theta\mybar\theta
(\mybar z_{3,\bfn} z_{i,\bfn}- z_{i,\bfn} \mybar z_{3,\bfn + \ih})\ ,\\
\mybar {\bfz}_{i,\bfn} &= \mybar z_{i,\bfn}
+\sqrt{2}\,\mybar\theta \epsilon_{ij}\xi_{j,\bfn} +\sqrt{2}\, \theta\mybar\theta
(\mybar z_{3,\bfn+\ih}\mybar z_{i,\bfn} -\mybar z_{i,\bfn}\mybar z_{3,\bfn})  \ , \\
%
{\bold \Xi_\bfn} &= \xi_{3,\bfn} +\sqrt{2}\, \theta
\Bigl(G_\bfn-\sqrt{2}\,\epsilon_{ij}\,\mybar z_{i,\bfn+\jh}\mybar
z_{j,\bfn}
\Bigr)  
-\sqrt{2}\, \theta \mybar\theta
(\mybar z_{3,\bfn + \xh + \yh}\xi_{3,\bfn} -\mybar z_{3,\bfn}\xi_{3,\bfn}
)\ ,\\
\mybar{\bold \Xi}_{\bfn} &= \chi_\bfn -\sqrt{2}\,\mybar\theta\Bigl(\mybar
G_\bfn-\sqrt{2}\,\epsilon_{ij}\, z_{i,\bfn}  z_{j,\bfn+\ih}
\Bigr) 
 +\sqrt{2}\, \theta \mybar\theta
(\mybar z_{3,\bfn}\chi_\bfn - \chi_\bfn\mybar z_{\bfn+\xh+\yh}) \ ,\\
{\bfs_\bfn} &= z_{3,\bfn} + \sqrt{2}\,\theta \psi_{3,\bfn} + \sqrt{2}\, \mybar\theta \lambda_\bfn
+\sqrt{2}\,\theta \mybar\theta \Bigl( 
\mybar z_{i,\bfn-\ih}
z_{i,\bfn-\ih} -z_{i,\bfn}\mybar z_{i,\bfn}
+id_{\bfn}\Bigr)\ ,\\
{\bold \Upsilon}_\bfn &=  \lambda_\bfn - 
\theta \Bigl(
\mybar z_{i,\bfn-\ih}  z_{i,\bfn-\ih} -
z_{i,\bfn}\mybar z_{i,\bfn} 
+[\mybar{z}_{3,\bfn}, z_{3,\bfn}] +i d_\bfn\Bigr) 
-\sqrt2 \theta \mybar{\theta}[\mybar{z}_{3,\bfn}, \lambda_\bfn]\ ,\\
\mybar {\bold\Upsilon}_\bfn& = \psi_{3,\bfn} + \mybar{\theta} \Bigl(
\mybar z_{i,\bfn-\ih}  z_{i,\bfn-\ih} - z_{i,\bfn}\mybar z_{i,\bfn}
-[\mybar{z}_{3,\bfn}, z_{3,\bfn}] +i d_\bfn\Bigr)
+ \sqrt2 \theta \mybar{\theta}[\mybar{z}_{3,\bfn}, \psi_{3,\bfn}]\ .
\end{aligned}
\eqn{sfields2d}\end{equation}
The orbifold projection of the  mother theory action
in \Eq{act8q2} may be written in terms of these lattice superfields as
\begin{equation}\begin{aligned}
S= \sum_{\bfn}\Tr &\left[\int d\theta d\mybar\theta\,  \left(
    \frac{1}{2}\mybar{\bold\Upsilon}_{\bfn}{\bold\Upsilon}_{\bfn} 
+\frac{1}{\sqrt{2}}{\bfs}_{\bfn}( {\bfz}_{i,\bfn}{{\mybar\bfz}}_{i,\bfn} - {{\mybar\bfz}}_{i,\bfn-
    \ih} {\bfz}_{i,\bfn-\ih})
-\frac{1}{2}\mybar{\bold\Xi}_{\bfn}
{\bold \Xi}_{\bfn}  \right) \right.\\ 
&\quad +  \left.
\int d\theta\,\Bigl(
\epsilon_{ij}\, {\bold \Xi}_{\bfn }\,{\bfz}_{i,\bfn}{\bfz}_{j,\bfn
    +\ih} 
\Bigr)  
- \int d\mybar\theta\,\Bigl( 
\epsilon_{ij}\,\mybar{\bold\Xi}_{\bfn
}\,\mybar {\bfz}_{i,\bfn+\jh}\mybar {\bfz}_{j,\bfn} 
\Bigr) 
\right]
\end{aligned}\eqn{lact2}\end{equation}

\subsection{The $C_{2v}$ and $U(1)^4$ symmetries of the $d=2$ lattice}
\label{sec:3b}

The action of the $C_{2v}$ and $U(1)^4$ lattice symmetries are
conveniently expressed 
in terms of superfields.  The generators of $C_{2v}$ are $C_2$,
corresponding to rotations of the lattice by $\pi$, and $\sigma_d$,
which reflects the lattice about the diagonal.  Their effect on the
various superfields of the theory are shown in
Table~\ref{tab:tab2}. Note that this symmetry does not commute with
supersymmetry, as the $C_2$ generator acts nontrivially on the
Grassmann parameter $\theta$. Also shown in Table~\ref{tab:tab2} are
the charges of the fields under 
the $U(1)^4$ symmetry.  We choose the four $U(1)$ charges to be $\bfr
= \{r_1,r_2\}$, $q_3$ and $q_4$;  the charges for the individual component
variables were given in Table~\ref{tab:tab1}. Note that both $q_3$ and
$q_4$ generate $R$-symmetries under which $\theta$ and $\mybar \theta$
are charged.  Furthermore, unlike in Minkowski space, one cannot find
a linear combination of $q_3$ and $q_4$ which is not an $R$-symmetry,
as $\theta$ and $\mybar \theta$ are independent in Euclidean
superspace. Also note that $q_4$ does not commute with the lattice
symmetry transformation $C_2$.  

 \setlength{\extrarowheight}{5pt}
\begin{table}[t]
\centerline{
\begin{tabular}
{|c||c|c|c|c|c|c|c|c|c|c|c|}
 \hline
&$  \theta $&$ \mybar \theta $&$ \bfz_{1,\bfn}$&$ \mybar \bfz_{1,\bfn}$ &$\bfz_{2,\bfn}$  &$\mybar \bfz_{2,\bfn}$
&$ \bfXi_\bfn$ &$\mybar\bfXi_\bfn$ & $\bfs_\bfn$ & $\bfU_\bfn$ & $ \mybar
\bfU_\bfn$\\ \hline
$C_2$ & -$\mybar\theta$ & -$\theta$& $\mybar \bfz_{1,-\bfn-\xh}$ & $\bfz_{1,-\bfn-\xh}$ & 
$\mybar \bfz_{2,-\bfn-\yh}$ & $\bfz_{2,-\bfn-\yh}$ &  
$\mybar \bfXi_{-\bfn}$ & $\bfXi_{-\bfn}$ & $\bfs_{-\bfn}$ &
$-\mybar\bfU_{-\bfn}$ & $-\bfU_{-\bfn}$\\
$\sigma_d$ &$ \theta$&$\mybar \theta$&
 $\bfz_{2,\tilde\bfn}$ &  $\mybar \bfz_{2,\tilde\bfn}$&  $\bfz_{1,\tilde\bfn}$ &
 $\mybar \bfz_{1,\tilde\bfn}$&
 $-\bfXi_{\tilde\bfn}$ &  $-\mybar \bfXi_{\tilde\bfn}$&
$\bfs_{\tilde \bfn}$ &  $\bfU_{\tilde \bfn}$ &  $ \mybar
\bfU_{\tilde \bfn}$ \\ \hline
$\bfr$ & 0 & 0 & $\{1,0\}$ & $\{-1,0\}$ & $\{0,1\}$ & $\{0,-1\}$ &
$\{-1,-1\}$ & $\{1,1\}$ & $\{0,0\}$ & $\{0,0\}$ & $\{0,0\}$ \\ \hline
$q_3$ & $\half$ & $\half$ & 0 & 0 & 0 & 0 & $\half$ & $\half$ & 1 & $\half$ & $\half$ \\ \hline
$q_4$ & $-\half$ & $\half$ & 0 & 0 & 0& 0 & $-\half$ & $\half$ & 0 & $-\half$ & $\half$ \\ \hline
 \end{tabular}}
\caption{\sl The action of the $C_2$ and $\sigma_d$ generators of the
  $C_{2v}$ lattice symmetry  on the superfields for the
  two-dimensional lattice of Fig.~1, as well as the $U(1)^4$ charges
  taken to be ${\bf r}=\{r_1,r_2\}$, $q_3$ and $q_4$.  The ${\bf r}$
  charges correspond to the position of the superfield in the unit
  cell of the lattice.
The coordinate  used are $\bfn=\{n_x,n_y\}$ and $\tilde \bfn=\{n_y,n_x\}$. Neither the $C_2$
nor the $q_4$ charge commute with supersymmetry.\label{tab:tab2}} 
\end{table}
 \setlength{\extrarowheight}{-2pt}

\subsection{The continuum limit of the $d=2$ lattice}
\label{sec:3c}

The lattice action we have defined has a large classical moduli space,
the space of possible values for our bosonic variables for which the
ground state energy is zero.  Following the procedure used in
Refs.~\cite{Kaplan:2002wv,Cohen:2003aa}, we now expand our lattice
action about the particular  point in moduli space
\begin{equation}
z_{1,\bfn} = z_{2,\bfn} = \frac{1}{\sqrt{2}\,a}\,{\bold 1}_k\ ,
\eqn{vev}\end{equation}
where ${\bold 1}_k$ is the $k\times k$ unit matrix, and $\rma$ is
interpreted as the lattice spacing.  We obtain the continuum
superfields by replacing the lattice coordinates $\bfn=\{n_x,n_y\}$with
continuous variables $\{x,y\}$,  and shifting the fields
\begin{equation}
\begin{aligned}
\bfPhi_1 &\equiv \bfz_1 - \frac{1}{\sqrt{2}\,\rma}\,{\bold 1}_k \ ,\qquad
&\frac{(\phi_1+iv_1)}{\sqrt{2}} &\equiv z_1 - \frac{1}{\sqrt{2}\,\rma}\,{\bold 1}_k \\
\bfPhi_2 &\equiv \bfz_2 - \frac{1}{\sqrt{2}\,\rma}\,{\bold 1}_k \ ,\qquad
&\frac{(\phi_2+iv_2)}{\sqrt{2}}&\equiv z_2 - \frac{1}{\sqrt{2}\,\rma}\,{\bold 1}_k\ ,
\end{aligned}
\end{equation}
defining the superfields $\bfPhi_i$ and the component fields $\phi_i$,
$v_i$ (not to be confused with the gauge fields of the mother
theory).  Both  $\phi_i$ and $v_i$ are Hermitean matrices.  We then expand both
the superfields in \Eq{sfields2d} and the lattice action \Eq{lact2}
in powers of the lattice spacing $\rma$ \footnote{Although we do not show
  it here, the free spectrum of the lattice action does
not have any fermion ``doublers'', so we are justified in keeping only
smooth fields in the continuum limit, ignoring states near the
edges of the Brillouin zone. How this works was  shown explicitly for the $\CQ=4$
theory of Ref.~\cite{Cohen:2003aa}}.  Defining the covariant
derivatives and field strength
\begin{equation}
D_1 = \partial_1 + i v_1\ ,\quad 
D_2 = \partial_2 + i v_2\ ,\quad 
v_{12} = -i[D_1,D_2]\ ,
\end{equation}
we find the continuum
superfields, up to terms of order $O(\rma)$:
\begin{equation}
\begin{aligned}
\bfPhi_m &= \frac{(\phi_m+iv_m)}{\sqrt{2}} + \sqrt{2}\theta \psi_m +
\theta\mybar\theta\left( D_m\mybar z +[\phi_m,\mybar
  z]\right)+O(\rma)\\
\mybar \bfPhi_m &= \frac{(\phi_m-iv_m)}{\sqrt{2}}
+\sqrt{2}\mybar\theta\,\epsilon_{mn}\xi_n + \theta\mybar\theta\left(D_m\mybar z -[\phi_m,\mybar
  z]\right)+O(\rma)\\
\bfs&=z + \sqrt{2}\,\theta\psi_3 +
\sqrt{2}\,\mybar\theta\lambda
+\sqrt{2}\theta\mybar\theta\left(-D_1\phi_1 - D_2 \phi_2 +
  id\right)+O(\rma)\\
\bfXi&= \xi_3 + \theta\left(\sqrt{2}\, G +
  (D_1\phi_2-D_2\phi_1-iv_{12} - [\phi_1,\phi_2])\right)
-\sqrt{2}\theta\mybar\theta[\mybar z,\xi_3]+O(\rma)\\
\mybar\bfXi&= \chi + \mybar\theta\left(-\sqrt{2}\,\mybar G +
  (D_1\phi_2-D_2\phi_1+iv_{12} + [\phi_1,\phi_2])\right)
+\sqrt{2}\theta\mybar\theta[\mybar z,\chi]+O(\rma)\\
\bfU &=\lambda + \theta\left(D_1\phi_1 + D_2\phi_2 - [\mybar z,z] -
  id\right) -\sqrt{2}\,\theta\mybar\theta [\mybar z,\lambda]+O(\rma)\\
\mybar\bfU&=\psi_3 - \mybar\theta\left(D_1\phi_1 + D_2\phi_2 + [\mybar z,z] -
  id\right) +\sqrt{2}\,\theta\mybar\theta [\mybar z,\psi_3]+O(\rma)\\
\end{aligned}
\eqn{csfieldsd2}
\end{equation}
From the definition of the continuum superfields and the gauge
transformation properties of the lattice variables, it is
straightforward to determine how the continuum superfields vary under smooth
gauge transformations. One finds that (up to $O(\rma)$ corrections) all
fields transform as $U(k)$ adjoints, except for the $\bfPhi_m$
fields which transform
inhomogeneously
\begin{equation}
\bfPhi_m \to U\bfPhi_m U^\dagger +\frac{1}{\sqrt{2}} U\partial_m
U^\dagger+O(\rma)\ , \qquad 
\mybar \bfPhi_m \to U\mybar \bfPhi_m U^\dagger -\frac{1}{\sqrt{2}} U\partial_m
U^\dagger+O(\rma)\ .
\end{equation}
It follows that we can define  the super-covariant derivatives
\begin{equation}
\CD_m = \partial_m + \sqrt{2} \bfPhi_m\ ,\qquad 
\mybar\CD_m = -\partial_m + \sqrt{2}\mybar \bfPhi_m\ ,
\end{equation}
which transform as $\CD_m\to U \CD_m U^\dagger +O(\rma)$, and similarly
for $\mybar \CD_m$.  From these one can construct chiral gauge field
strength superfields $\CV$ and $\mybar \CV$
\begin{equation}
\begin{aligned}
\CV_{mn} &= -i[\CD_m,\CD_n]\\ &= 
\left(v_{mn}   -i\left(D_m\phi_n-D_n\phi_m\right)-i[\phi_m,\phi_n]\right)\\
&\qquad-2i\,\theta\left(D_m\psi_n-D_n\psi_m+[\phi_m,\psi_n]-[\phi_n,\psi_m]\right)\\
&
\qquad +\sqrt{2}\,
\theta\mybar\theta [v_{mn} + i\left(D_m\phi_n-D_n\phi_m\right),\mybar
z]\ ,\\&\\
\mybar\CV_{mn}&=-i[\mybar\CD_m,\mybar\CD_n]\\ 
&=  \left(v_{mn}
  +i\left(D_m\phi_n-D_n\phi_m\right)-i[\phi_m,\phi_n]\right)\\
&\qquad+2i\,\mybar\theta\left(D_m\epsilon_{np}\xi_p-D_n\epsilon_{mp}\xi_p-\epsilon_{np}[\phi_m,\xi_p]
+\epsilon_{mp}[\phi_n,\psi_p]\right)\\
&\qquad -\sqrt{2}\,
\theta\mybar\theta [v_{mn} + i\left(D_m\phi_n-D_n\phi_m\right),\mybar
z]\ , 
\end{aligned}
\end{equation}
as well as a ``vector''
gauge superfield strength $\CW$:
\begin{equation}
\begin{aligned}
\CW &= \sum_{m=1}^2\,[\CD_m,\mybar\CD_m]= \sqrt{2}\,\partial_m(\bfPhi_m
+\mybar\bfPhi_m) +2 [\bfPhi_m,\mybar\bfPhi_m]\\
&= 2 D_m\phi_m + 2 \theta(D_m\psi_m - [\phi_m,\psi_m]) +
 2 \mybar\theta \epsilon_{mn} (D_m\xi_n +[\phi_m,\xi_n])
 \\
&\qquad + 2\sqrt{2}\,\theta\mybar\theta\left(D_mD_m\mybar z -
  [\phi_m,[\phi_m,\mybar z]]-\sqrt{2}\,\{\psi_m,\epsilon_{mn}\xi_n\}\right)\ .
\end{aligned}
\end{equation}
The action \Eq{lact2} may be compactly expressed in terms of these
fields in a manifestly gauge and $\CQ=2$ supersymmetric way as
\begin{equation}
S = \frac{1}{2g_2^2}\,\int d^2x\, \Tr 
\left[ \int d\theta\,
d\mybar\theta\,\left(\mybar \bfU \bfU + 
\frac{1}{\sqrt{2}}\,\bfs\CW -
\mybar\bfXi \bfXi \right)
-i\,\int d\mybar\theta\  \mybar\bfXi \,\mybar\CV_{12} + i\,\int
  d\theta\ \bfXi \,\CV_{12}\right] + O(\rma)\ ,
\end{equation}
with $g_2^2 \equiv g^2 \rma^2$. After some algebra it is possible to show
that the above action in the $\rma\to 0$ limit is
identical to the action of the target theory (reproduced here from \Eq{targ2})
\begin{equation}
S= \frac{1}{g_2^2} \int d^2x\,  \Tr\Biggl[\frac{1}{4} v_{mn} v_{mn}
+\frac{1}{2} (D_m s_a)^2 +\mybar\Psi_i \gamma_m D_m
\Psi_i + \mybar \Psi_i [s_0,\,\Psi_i] + i\mybar \Psi_i \gamma_3
\bftau_{ij}\cdot [{\boldsymbol s},\,\Psi_j]
-\fourth[s_a,\,s_b]^2\Biggr]\\
\eqn{targ2b}\end{equation}
with the substitutions
\begin{equation}
\Psi_1 =
\begin{pmatrix}
\xi_1\\ \xi_2 \\
\end{pmatrix}\ , \qquad
\Psi_2 =
\begin{pmatrix}
\ \lambda\\ -\xi_3 \\
\end{pmatrix} \ ,\qquad
\mybar\Psi_1 =
\begin{pmatrix}
-\chi & \psi_3 \\
\end{pmatrix} \ ,\qquad
\mybar\Psi_2 =
\begin{pmatrix}
\psi_1 & \psi_2
\end{pmatrix} \ .
%
\eqn{ferms2}\end{equation}

\begin{equation}
\begin{pmatrix}
s_0\\
s_1\\
s_2\\
s_3
\end{pmatrix} = 
\begin{pmatrix}
\phi_1 \\
-(z+\mybar z)/\sqrt{2}\\
i(z-\mybar z)/\sqrt{2}\\
\phi_2
\end{pmatrix}
\end{equation}
in the following $\gamma$-matrix basis:
\begin{equation}
\gamma_1 = -\sigma_3\ ,\qquad \gamma_2 = -\sigma_1\ ,\qquad \gamma_3 =
\sigma_2\ .
\end{equation}
It is not perversity that leads us to choose an off-diagonal matrix
for the chirality matrix $\gamma_3$---since chiral rotations are
anomalous in the continuum theory, they mix fermion components from
different locations on the lattice. As we chose a basis in \Eq{ferms}
where each fermion component of the continuum  corresponds to a
particular lattice variable, the $\gamma_3$ matrix must
necessarily be off-diagonal.

\subsection{Renormalization on the $d=2$ lattice}
\label{sec:3d}

The discussion of renormalization for this two dimensional lattice is
similar to those given in Refs.~\cite{Kaplan:2002wv,Cohen:2003aa}. We
have shown that the tree level action has the desired continuum limit;
we must now argue that there are no relevant or marginal
operators generated radiatively which violate the
Lorentz and $\CQ=8$ supersymmetry of the target theory.  An essential
part of our argument relies on the fact that any such operators must
be invariant under the exact $\CQ=2$ supersymmetry, $U(k)$ gauge
symmetry and the $C_{2v}$ and $U(1)^4$ global symmetries respected by
the lattice action.

Consider the addition to the action of chiral operators
$\CO$ and $\mybar \CO$, or a vector operator $\widehat \CO$: 
\begin{equation}
  \delta S=\frac{1}{g_2^2} \int\! d^2\!x\ \left[\left( \int\! d\theta\
      C\,\CO\, +  \int\! d\mybar\theta\ 
    \mybar C\,\mybar\CO\right) + \int\! d\theta\, d\mybar\theta\ 
 \widehat C \widehat \CO\right]\ .
\end{equation}
We will ignore the anti-chiral operator $\mybar \CO$, which has the
same power counting as the chiral operator $\CO$ and is related by the
$C_2$ lattice symmetry.  For power counting purposes,
$\frac{1}{g_2^2}\int\! d^2\!x$ has scaling dimension $-4$, $d\theta$,
$d\mybar\theta$, $\CD$ and $\mybar\CD$ all scale with dimension
$+\half$.  A contribution to any operator coefficient at $\ell$ loops
is proportional to $g_2^{2\ell}$.  Since the coupling $g_2$ has mass
dimension 1, the loop expansion is an expansion in
powers of the dimensionless parameter $(g_2^2 \rma^2)^\ell$.
Consequently the operator coefficients have a loop expansion of the
form
\begin{equation}
\begin{aligned}
  C &= \rma^{p-7/2}\,\sum_\ell c_\ell\, (g_2^2 \rma^2)^\ell\ ,\\
\widehat  C &= \rma^{\widehat p-3}\,\sum_\ell \widehat c_\ell\, (g_2^2 \rma^2)^\ell\ ,
\end{aligned}
\end{equation}
where the chiral and vector operators $\CO$ and $\widehat\CO$ are assumed
to have mass dimension $p$ and $\widehat p$ respectively.
 The dimensionless expansion coefficients $c_\ell$ and $\widehat c_\ell$ depend at most
logarithmically on the lattice spacing $\rma$. Considering $\ell\ge 1$,
we see that the dangerous operators are the ones satisfying
\begin{equation}
p-\frac{3}{2} \le 0\ ,\qquad
\widehat p -1 \le 0\ .
\end{equation}

The only possibilities for $\CO$ on dimensional grounds are $\CO=\Tr
\bfXi$ or $\CO=\Tr \bfU$; however the former is excluded by its
oddness under the $\sigma_{d}$ generator of the lattice $C_{2v}$
symmetry (see Table~\ref{tab:tab1}), while the $\theta$ component of
$\Tr \bfU$ is a total derivative and does not contribute to the
action.  As for the vector superfield operator operator, the only
option is $\widehat\CO=\Tr S$.  This contributes a tadpole for the
$U(1)$ $d$-term, and so looks like a standard Fayet-Illiopoulos term,
which in this case contributes only to a harmless  cosmological constant.  Thus the $\CQ=8$ target
theory is obtained without any fine tuning of parameters, since no
marginal or relevant operators spoil its emergence in the $\rma\to 0$
limit.

\subsection{Fixing the moduli}
\label{sec:3e}

The last loose end to tie up is the  fixing of moduli
(bosonic zeromodes).  The analysis is identical to that in \S6 of
Ref.\cite{Cohen:2003aa}, and so we only recapitulate the results here.
The first point is that fixing the ``vacuum expectation values'' of
the $z_{1,\bfn}$ and $z_{2,\bfn}$ link fields as in \Eq{vev} requires the
addition of a supersymmetry breaking term to the action which lifts
the degeneracy of the moduli.  In terms of the continuum variables it
takes the form of a small mass term $\mu$ for the $s_a$ boson quartet,  which vanishes in
the large volume limit.  In terms of the lattice variables, such a
term can take the form of the last line in \Eq{targ2}:
\begin{equation}
\frac{a^2\mu^2}{2}\left[\left( z_{1,\bfn}\mybar z_{1,\bfn}
    -\frac{1}{2a^2}\right)^2 + \left( z_{2,\bfn}\mybar
    z_{2,\bfn} -\frac{1}{2a^2}\right)^2+2\frac{\mybar z_{3,\bfn}
    z_{3,\bfn}}{a^2}\right]\ .
\eqn{ssb}\end{equation}
The parameter $\mu$ can be taken to be $\lesssim 1/L = 1/N\rma$.
The mass terms in \Eq{ssb} serve to fix the scalar zeromodes, just as
an external magnetic field whose strength scales to zero in the large
volume limit can be used to study systems with spontaneous
magnetization. There is the separate issue of the infrared divergences
of nonzero momentum modes of the moduli, which one might expect to be
severe in two dimensions, spoiling our expansion about $\langle z_{1,\bfn}
\rangle = \langle z_{2,\bfn} \rangle = \frac{1}{2\sqrt{\rma}}$;  however, as
shown in Ref.~\cite{Cohen:2003aa} this is in fact not a problem with
our method, provided the continuum and large volume limits are taken so
that $\rma^2 g_2^2 \ln N = \rma^4 g^2 \ln N\to 0$.

\section{The three dimensional lattice}
\label{sec:4}

\subsection{The $d=3$ lattice action and its symmetries}
\label{sec:4a}

The three dimensional lattice is derived from
the $\CQ=8$ mother theory by performing a $Z_N^3$ orbifold projection,
where the $Z_N^3$ symmetry is defined by the charges $r_{1,2,3}$ in
Table~\ref{tab:tab1}. This lattice will describe the $\CN=4$ supersymmetry
in three dimensions in the continuum, whose action is given
\Eq{targ3}.
The lattice we obtain takes the form  shown in
Fig.~\ref{fig:fig2}, with $ \lambda$ residing at
the sites;  $\{z_a,\,\mybar z_a,\, \psi_a\}$ on the $\hat {\bf z}_a$-links
for $a=1,2,3$; the $\xi_a$ on the diagonal face links; and $\chi$ on
the superdiagonal link. 
This lattice possesses a $C_{3v}\cong S_3$ point group symmetry,
consisting of $2\pi/3$ rotations around the $\chi$ link, as well as
reflections about the three planes containing both $\chi$ and $\psi_i$
links. The lattice action is also invariant under a single
supersymmetry transformation, obtained by setting $\kappa$ in the
supersymmetry transformations of the mother theory \Eq{dq6} to
\begin{equation}
\kappa =\begin{pmatrix}\ \eta\  & \ 0\  \\ \ 0\  & \ 0\  \\ \ 0\  &
  \ 0\  \\ \ 0\  & \ 0\  \end{pmatrix}\ .
\end{equation} 

Following the procedure followed for the two dimensional lattice, we
introduce a Grassmann coordinate $\theta$ and define the supersymmetry
transformation to be
\begin{equation}
\delta = i\eta \,Q\ ,\qquad Q = \frac{\partial\ }{\partial \theta}\ ,
\end{equation}
which acts on the superfields 
\begin{equation}
\begin{aligned}
{\bold\Lambda}_{\bfn}&= \lambda_{\bfn} -\theta\, \left(
  \mybar z_{a,\bfn-\ah}\,
  z_{a,\bfn-\ah}-  z_{a,\bfn}\,\mybar
  z_{a,\bfn}+ i d_{\bfn}\right)\ ,\\
 &&\\
{\bfz}_{a,\bfn} &= z_{a,\bfn} + \sqrt{2}\,\theta \,\psi_{a,\bfn}\ ,\\
 &&\\
{\bold\Xi}_{a,\bfn}&= \xi_{a,\bfn} -  2\theta\,\epsilon_{abc}\,\mybar
z_{b,\bfn+\ch} \mybar
z_{c,\bfn}\ .
\end{aligned}
\end{equation}
In the above equation  repeated  indices are summed over $1,2,3$, and  $\ah$ is a unit lattice
vector in the positive $z_a$ direction. In terms of individual
components, the supersymmetry transformations 
are given by 
\begin{equation}
\begin{aligned}
\delta z_{a,\bfn} &= i\sqrt{2}\,\eta\psi_{a,\bfn} \\
\delta \mybar z_{a,\bfn} &= 0\\
\delta\psi_{a,\bfn}&=0 \\
\delta\xi_{a,\bfn} &=-2i \eta\epsilon_{abc}\left(\mybar
  z_{b,\bfn+\boldsymbol{\ch}}\mybar z_{c,\bfn}\right)\\
\delta\lambda_\bfn &= -i \eta
  \left(\mybar z_{a,\bfn-\ah}\,
  z_{a,\bfn-\ah}-  z_{a,\bfn}\,\mybar
  z_{a,\bfn} +i d_{\bfn}\right) \\ 
\delta\chi_\bfn&=0\\
\delta d_\bfn &= -\sqrt{2}\,\eta \left(\mybar
  z_{a,\bfn-\ah} \psi_{a,\bfn-\ah}-\psi_{a,\bfn}\mybar
  z_{a,\bfn}\right)\ .
\eqn{dfieldsb}\end{aligned}
\end{equation} 
  Note that $\mybar z_a$, $\psi_a$ and $\chi$ are all
supersymmetric singlets, as is the $\theta$ component of any
superfield.

The lattice action we obtain may be written in manifestly $\CQ=1$
supersymmetric form as 
\begin{equation}
\begin{aligned}
S = \frac{1}{g^2} \sum_{\bfn}\Tr&\biggl(\int d\theta\,\Bigl[ -\half \bfl_\bfn
  \partial_\theta \bfl_\bfn -  \bfl_\bfn\left(\mybar
    z_{a,\bfn-\ah}
    \bfz_{a,\bfn-\ah} - \bfz_{a,\bfn} \mybar
    z_{a,\bfn}\right)
+ \epsilon_{abc}\, \bfXi_{a,\bfn}\bfz_{b,\bfn}\bfz_{c,\bfn
  +\bh} 
\Bigr]\\
&-\sqrt{2}\,\chi_\bfn\left(\mybar z_{a,\bfn +\boldsymbol{\omega}_a}
  \bfXi_{a,\bfn} - \bfXi_{a,\bfn+\ah}\mybar
  z_{a,\bfn}\right)\biggr)\ ,\\
\eqn{ssact3}\end{aligned}
\end{equation} 
with
\begin{equation}
\boldsymbol{\omega}_a\equiv \sum_{b\ne a}\bh =
\{1,1,1\} - \ah\ .
\end{equation}
The last term in the action is not integrated over $\theta$, even
though it contains the nontrivial superfield $\bfXi$. However, due to
the $\epsilon_{abc}$ tensor in the $\theta$ component of $\bfXi$, one
can see that the $\theta$ component of the $\Tr \chi[\mybar z , \bfXi]$ operator in the action
above identically vanishes, so that $\Tr \chi_\bfn\left(\mybar z_{a,\bfn +\boldsymbol{\omega}_a}
  \bfXi_{a,\bfn} - \bfXi_{a,\bfn+\ah}\mybar
  z_{a,\bfn}\right)$  is $\theta$-independent and hence
supersymmetric.

The transformations of the superfields under the global symmetries of
the theory are given in Table~\ref{tab:tab3}. In this table the $S_3$
generators $R$
and $\Sigma$ are given by
\begin{equation}
R = 
\begin{pmatrix} 0 & 1& 0\\ 
0 & 0& 1\\
1 &0 &0 \\ \end{pmatrix}\qquad
\Sigma = \begin{pmatrix} 0 & 1& 0\\1 & 0& 0\\ 0 &0 &1 \\ \end{pmatrix}
\ .
\end{equation}

 \setlength{\extrarowheight}{5pt}
\begin{table}[t]
\centerline{
\begin{tabular}
{|c||c|c|c|c|c|c|}
 \hline
&$  \theta $&$ \bfz_{a,\bfn}$&$ \mybar z_{a,\bfn}$ &$ \bfXi_{a,\bfn}$
&$ \Lambda_\bfn$ & $\chi_{\bfn}$\\ \hline \hline
$C_3$ & $\theta$ & $R_{ab}\,\bfz_{b,R\bfn}$ & $R_{ab}\,\mybar z_{b,R\bfn}$
& $R_{ab}\, \bfXi_{b,R\bfn}$ & $ \Lambda_\bfn$ & $\chi_{\bfn}$\\ \hline
$\sigma_3$ &$\theta$ &  $\Sigma_{ab}\,\bfz_{b,\Sigma\bfn}$ & $\Sigma_{ab}\,\mybar z_{b,\Sigma\bfn}$
& $-\Sigma_{ab}\, \bfXi_{b,\Sigma\bfn}$ & $ \Lambda_\bfn$ & $\chi_{\bfn}$\\ \hline
$\bfr$ & ${\bold {0}} $ &  ${\bold{ \hat e}}_a$ & - ${\bold {\hat e}}_a$
&${\bold{ \hat e}}_a -\sum_b {\bold {\hat e}}_b$ & ${\bold {0}}$ & $\sum_b
{\bold {\hat e}}_b$\\ \hline
$q_4$ & $-\half $ & $0$ & $0$ & $-\half$ & $-\half$ & $\half$ \\ \hline
 \end{tabular}}
\caption{\sl The transformation properties of fields on the three
  dimensional lattice under the global $S_3\times U(1)^4$ symmetry.  $C_3$ and $\sigma_3$
  generate the $S_3$ group, with the former consisting of $2\pi/3$
  rotations about the vector $\{1,1,1\}$, and $\sigma_3$
  corresponding to reflections about the plane containing the
  $\{0,0,1\}$ and $\{1,1,1\}$ vectors. The matrices $R$ and $\Sigma$
  are given in the text.  The global $U(1)^4$ symmetry is taken to be
  generated by the three ${\bf r}$ charges, and $q_4$, the latter
  being an $R$-charge on the lattice.\label{tab:tab3}} 
\end{table}
\bigskip

\subsection{The continuum limit of the $d=3$ lattice}
\label{sec:4b}

As we did for the two dimensional lattice, we now reexpress the three
dimensional lattice theory in terms of continuum superfields which
facilitates the analysis of its properties under renormalization.  
We define the shifted link fields
\begin{equation}
\frac{(\phi_a+iv_a)}{\sqrt{2}}
 \equiv 
z_a - \frac{1}{\sqrt{2}\,\rma}\,{\bold
  1}_k\ ,\qquad
\bfPhi_a  \equiv\bfz_a - \frac{1}{\sqrt{2}\,\rma}
=\frac{(\phi_a+iv_a)}{\sqrt{2}} +\sqrt{2}\theta \psi_a\ ,
\end{equation}
where the subscript $a$ runs over $1,2,3$.  The ordinary gauge
covariant derivatives and field strength are defined as
\begin{equation}
D_m = \partial_m + i v_m\ ,\quad 
v_{mn} = -i[D_m,D_n]\ .
\end{equation}
where the spacetime indices $m,n=1,2,3$.
The continuum limit of the lattice fields involves keeping only the
smooth configurations, as explicit computation shows that the
perturbative propagators exhibit no poles near the edges of the
Brillouin zone.  We can write the continuum superfields (up to $O(\rma)$ corrections) as
\begin{equation}
\begin{aligned}
\bfPhi_a &= \frac{\phi_a + i v_a}{\sqrt{2}} +\sqrt{2}\,\theta\,\psi_a\ ,\\
&\\
\bfXi_a &= \xi_a + \half \theta\,\epsilon_{abc} \left(D_b\phi_c-D_c\phi_b -
  [\phi_b,\phi_c]-iv_{bc}\right) + O(\rma)\ ,\\ &\\
\bfl&= \lambda+\theta\left(D_a\phi_a-id\right) + O(\rma)\ ,
\end{aligned}
\end{equation}
where each field transforms as  an adjoint under the  $U(k)$ gauge
symmetry, and is a function of the three coordinates in our three
dimensional Euclidean spacetime.  The lattice action is again most
easily written by introducing super-covariant derivatives, which in
the present case are
\begin{equation}
\CD_m \equiv \partial_m + \sqrt{2}\, \Phi_m\ ,\qquad \mybar \CD_m \equiv
-\partial_m + (\phi_m -iv_m) = -D_m + \phi_m\ .
\end{equation}
From these one can construct the field strengths
\begin{equation}
\begin{aligned}
\CV_{mn} &= -i[\CD_m,\CD_n]\\  &= -i\left(D_m\phi_n-D_n\phi_n +
  [\phi_m,\phi_n]+iv_{mn}\right) -2i\theta\left(D_m\psi_n-D_n\psi_m +
  [\psi_m,\phi_n]-[\psi_n,\phi_m]\right)\ ,\\&\\ 
\CW &= [\CD_m,\mybar\CD_m]= 2D_m\phi_m +
  2\theta(D_m\psi_m + [\psi_m,\phi_m])\ ,
\end{aligned}
\end{equation}
where $m$ is summed over $1,2,3$ in the expression for $\CW$.  The
lattice action \Eq{ssact3} may then be rewritten in terms of continuum
fields, with $g_3^2 \equiv \rma^3 g^2$,  as
\begin{equation}
S =\frac{1}{g_3^2} \int d^3x\, \Tr\left( -\chi[\mybar\CD_a,\bfXi_a]+\int d\theta\,
    \left[-\half \Lambda \partial_\theta \Lambda +\half \Lambda \CW +
      \frac{i}{4} \epsilon_{abc}\bfXi_a \CV_{bc}\right]\right)
  +O(\rma)\ .
\eqn{s3oa} \end{equation}
In terms of components, after eliminating the auxiliary field $d$, this yields the desired target theory
\begin{equation}
S = \frac{1}{g_3^2}\int d^3 x\,\Tr\Biggl[\frac{1}{4} v_{mn} v_{mn}
+\frac{1}{2} (D_m \phi_a)^2 +\mybar\Psi_i \sigma_m D_m
\Psi_i  - \mybar \Psi_i 
\tau^a_{ij}\cdot [{ \phi_a},\,\Psi_j]
-\frac{1}{4}[\phi_a,\,\phi_b]^2\Biggr]\ .\\
\eqn{targ3b}\end{equation}
where
 the fermions fields are given by
\begin{alignat}{2}
\Psi_1 &= \frac{1}{\sqrt 2}
\begin{pmatrix}
\psi_3-i\chi \cr
 \psi_1+i\psi_2  \\
\end{pmatrix}
\ ,&\qquad
\mybar \Psi_1 &=\frac{i}{\sqrt 2} 
\begin{pmatrix}
\xi_3+i\lambda \cr
 \xi_1-i\xi_2  \\
\end{pmatrix}^T \,\
\\ &&&\\
\Psi_2 &= \frac{1}{\sqrt 2}
\begin{pmatrix}
-\psi_1 + i\psi_2 \cr 
\psi_3+i\chi
\end{pmatrix}\ ,&\qquad
\mybar \Psi_2 &= \frac{i}{\sqrt 2}
\begin{pmatrix}
-\xi_1-i\xi_2  \cr
\xi_3-i\lambda \\
\end{pmatrix}^T\,
\end{alignat}
in a basis where the three gamma matrices are just the Pauli matrices,
\beq
\gamma_m = \sigma_m\ .
\eeq

\subsection{Renormalization on the $d=3$ lattice}
\label{sec:4c}
To find out whether operators might be radiatively induced which
could spoil the continuum limit, we follow the same procedure as in
\S~\ref{sec:3d}. Counterterms in the action will take the generic form
\footnote{One might wonder if there could be operators
 induced which do not have to be integrated over $\theta$, such as the
 first one in \Eq{s3oa}.  However, it is not hard to show that there
 are no such operators that satisfy our criteria for ``dangerous
 operators'', due to the combined strictures of gauge invariance and
 the $S_3$ symmetry.}
\begin{equation}
\delta S = \frac{1}{g_3^2} \int d^3x\,\left[ \int d\theta\,
  C\CO\right]\ ,
\end{equation}
where $\CO$ is an operator of dimension $p$.  The coupling $g_3^2$
scales like mass, while $d^3x$ and $d\theta$ have mass dimension $(-3)$
and $1/2$ respectively. Then on dimensional grounds, contributions to
$C$  in a loop expansion are of  the form
\begin{equation}
C = \rma^{p-7/2} \sum_\ell c_\ell (g_3^2 \rma)^\ell
\end{equation}
where $\ell$ counts the number of loops and $c_\ell$ is dimensionless,
depending at most logarithmically on $\rma$.  If $\CO$ is an operator
which violates the symmetries of the target theory then its coefficient
should vanish in the $\rma\to 0$ limit, or the continuum limit will be
spoiled.  Since radiative corrections begin at $\ell=1$, it follows
that we need to check whether the theory allows  operators with $p\le
\frac{5}{2}$ which respect the exact symmetries of the lattice.  

The operator $\CO$ must be Grassmann, and hence its dimension must be
half integer.  It  must have $U(1)$ charge $q_4=
-\half$, and be invariant under the gauge and $S_3$ symmetries.
Denoting generic bosonic superfields as $B$ and fermion superfields as
$F$, we find that there is no operator of dimension $p\le \half$; at
dimensions $p=\frac{3}{2}$  $\CO$ must take the form $\partial_\theta
B$ or $F$.  The former is a total derivative, but the latter is a
possibility.  However when the symmetries are taken into account, the
only possible operator one could add is $\CO=\Tr \Lambda$.  This is a
Fayet-Iliopoulos term, and as discussed previously, it only
contributes to a cosmological constant, and does not affect the
excitations of theory.

At dimension $p=\frac{5}{2}$  the choices for $\CO$ are
$B\partial_\theta B'$ and $F B$.  The former can be ruled out as 
there is no operator of that form with $q_4=-\half$.  However there are
two similar  operators of the form $F B$ which are invariant under all the
symmetries, namely
\begin{equation}
\CO_1 = \frac{1}{\sqrt{2}}\sum_a \Tr \Lambda\left(\bfPhi_a + \mybar \phi_a\right)\ ,\quad
{\rm and}\quad
\CO_2 = \frac{1}{\sqrt{2}}\sum_a \Tr \Lambda\, \Tr\left(\bfPhi_a + \mybar \phi_a\right)\ .
\end{equation}
Note that $\Phi_a \text{ and } \bar
 \phi_a$ must appear in this combination for the
inhomogeneous shift under gauge transformations to cancel.  In component form,
these operators are equivalent to
$ \sum_a
\Tr\left(-\lambda \psi_a + (D_b\phi_b-id) \phi_a\right)$ and 
its double trace analogue.
On the elimination of $d$, such operators would induce the scalar
mass term $\Tr \left(\sum_a \phi_a\right)^2$.
 The  coefficients of these operators   can have at most a logarithmic divergence at
one loop, and vanishing contributions at higher loops.  Therefore,
while we cannot rule out that these operators could be radiatively
induced, in principle the critical couplings  can be determined from
a one-loop calculation on the lattice.

The discussion about how to fix the moduli of the three dimensional
theory is similar to that given in \S~\ref{sec:3e} for the two
dimensional case, and we will not repeat it here.

In conclusion, the three dimensional lattice allows two
supersymmetric counterterms which could be radiatively induced at one
loop, and which would spoil the continuum limit.  The coefficient of
this operator only runs logarithmically with scale, and it must be
tuned to the critical point, either numerically, or theoretically by
performing a one loop computation on the lattice.

\section{Discussion}
\label{sec:5}
We have explicitly constructed a nonperturbative regulator for quantum field
 theories with eight supercharges in two and three Euclidean
 dimensions. Although there may be challenges to overcome, such as a
 potential sign problem with the fermion determinant\footnote{J. Giedt,
 private communication} (as there is in the two dimensional theory with
 four supercharges\cite{Giedt:2003ve}), we hope that eventually these lattice may
 be of use for numerical simulations. While there is a substantial
 literature on these theories already, most previous theoretical
 investigations 
 have centered on trying to understand the structure of their
 moduli spaces.  Presumably numerical simulations would begin by
 establishing that the supersymmetric Ward identities are satisfied,
 and proceed with a study of the particle spectrum, about which
 little is known.  
 
We are  optimistic that these lattices could prove useful for
 theoretical investigations as well, such as furthering our
 understanding of mirror symmetry in three
 dimensions\cite{Kapustin:1999ha}, or leading to the construction of a nonperturbative
 Nicolai map\cite{Nicolai:1980jc} in a fully regulated supersymmetric
 gauge theory. 

A final paper in this series in preparation will address 
 the construction
 of spacetime lattices for theories with sixteen supercharges.


\acknowledgments
D.B.K. and M.U. were supported in part by DOE grant DE-FGO3-00ER41132,
E.K by DOE grant DE-FG03-96ER40956, and A.G.C. by DOE grant
DE-FG03-91ER-40676. 

\bibliography{latticeSUSY2}

\providecommand{\href}[2]{#2}\begingroup\raggedright\begin{thebibliography}{10}

\bibitem{Kaplan:2002wv}
D.~B. Kaplan, E.~Katz, and M.~Unsal, {\it Supersymmetry on a spatial lattice},
  {\em JHEP} {\bf 05} (2003) 037,
  [\href{http://xxx.lanl.gov/abs/hep-lat/0206019}{{\tt hep-lat/0206019}}].

\bibitem{Cohen:2003aa}
A.~G. Cohen, D.~B. Kaplan, E.~Katz, and M.~Unsal, {\it Supersymmetry on a
  euclidean spacetime lattice I: A target theory with four supercharges},
  \href{http://xxx.lanl.gov/abs/http://arXiv.org/abs/hep-lat/0302017}{{\tt
  http://arXiv.org/abs/hep-lat/0302017}}.

\bibitem{Douglas:1996sw}
M.~R. Douglas and G.~W. Moore, {\it D-branes, quivers, and ale instantons},
  \href{http://xxx.lanl.gov/abs/http://arXiv.org/abs/hep-th/9603167}{{\tt
  http://arXiv.org/abs/hep-th/9603167}}.

\bibitem{Arkani-Hamed:2001ca}
N.~Arkani-Hamed, A.~G. Cohen, and H.~Georgi, {\it (De)constructing dimensions},
   {\em Phys. Rev. Lett.} {\bf 86} (2001) 4757--4761,
  [\href{http://xxx.lanl.gov/abs/http://arXiv.org/abs/hep-th/0104005}{{\tt
  http://arXiv.org/abs/hep-th/0104005}}].

\bibitem{Arkani-Hamed:2001ie}
N.~Arkani-Hamed, A.~G. Cohen, D.~B. Kaplan, A.~Karch, and L.~Motl, {\it
  Deconstructing (2,0) and little string theories},
  \href{http://xxx.lanl.gov/abs/http://arXiv.org/abs/hep-th/0110146}{{\tt
  http://arXiv.org/abs/hep-th/0110146}}.

\bibitem{Kaplan:1992bt}
D.~B. Kaplan, {\it A method for simulating chiral fermions on the lattice},
  {\em Phys. Lett.} {\bf B288} (1992) 342--347,
  [\href{http://xxx.lanl.gov/abs/http://arXiv.org/abs/hep-lat/9206013}{{\tt
  http://arXiv.org/abs/hep-lat/9206013}}].

\bibitem{Narayanan:1995gw}
R.~Narayanan and H.~Neuberger, {\it A construction of lattice chiral gauge
  theories},  {\em Nucl. Phys.} {\bf B443} (1995) 305--385,
  [\href{http://xxx.lanl.gov/abs/http://arXiv.org/abs/hep-th/9411108}{{\tt
  http://arXiv.org/abs/hep-th/9411108}}].

\bibitem{Neuberger:1998fp}
H.~Neuberger, {\it Exactly massless quarks on the lattice},  {\em Phys. Lett.}
  {\bf B417} (1998) 141--144,
  [\href{http://xxx.lanl.gov/abs/http://arXiv.org/abs/hep-lat/9707022}{{\tt
  http://arXiv.org/abs/hep-lat/9707022}}].

\bibitem{Giedt:2003ve}
J.~Giedt, {\it Non-positive fermion determinants in lattice supersymmetry},
  \href{http://xxx.lanl.gov/abs/hep-lat/0304006}{{\tt hep-lat/0304006}}.

\bibitem{Diaconescu:1997gu}
D.-E. Diaconescu and N.~Seiberg, {\it The coulomb branch of (4,4)
  supersymmetric field theories in two dimensions},  {\em JHEP} {\bf 07} (1997)
  001, [\href{http://xxx.lanl.gov/abs/hep-th/9707158}{{\tt hep-th/9707158}}].

\bibitem{Witten:1997yu}
E.~Witten, {\it On the conformal field theory of the higgs branch},  {\em JHEP}
  {\bf 07} (1997) 003, [\href{http://xxx.lanl.gov/abs/hep-th/9707093}{{\tt
  hep-th/9707093}}].

\bibitem{Seiberg:1996bs}
N.~Seiberg, {\it IR dynamics on branes and space-time geometry},  {\em Phys.
  Lett.} {\bf B384} (1996) 81--85,
  [\href{http://xxx.lanl.gov/abs/hep-th/9606017}{{\tt hep-th/9606017}}].

\bibitem{Seiberg:1996nz}
N.~Seiberg and E.~Witten, {\it Gauge dynamics and compactification to three
  dimensions},  \href{http://xxx.lanl.gov/abs/hep-th/9607163}{{\tt
  hep-th/9607163}}.

\bibitem{Chalmers:1997xh}
G.~Chalmers and A.~Hanany, {\it Three dimensional gauge theories and
  monopoles},  {\em Nucl. Phys.} {\bf B489} (1997) 223--244,
  [\href{http://xxx.lanl.gov/abs/hep-th/9608105}{{\tt hep-th/9608105}}].

\bibitem{Kapustin:1999ha}
A.~Kapustin and M.~J. Strassler, {\it On mirror symmetry in three dimensional
  abelian gauge theories},  {\em JHEP} {\bf 04} (1999) 021,
  [\href{http://xxx.lanl.gov/abs/hep-th/9902033}{{\tt hep-th/9902033}}].

\bibitem{Nicolai:1980jc}
H.~Nicolai, {\it Supersymmetry and functional integration measures},  {\em
  Nucl. Phys.} {\bf B176} (1980) 419--428.

\end{thebibliography}\endgroup
\bibliographystyle{JHEP} 

\end{document}